\documentclass[amsmath,amssymb,nofootinbib,notitlepage,prd,superscriptaddress,twocolumn]{revtex4-2}
\usepackage[T1]{fontenc}
\usepackage[usenames,dvipsnames]{xcolor}
\usepackage{graphicx,hyperref,orcidlink,float}
\hypersetup{colorlinks=true,citecolor=RoyalBlue,linkcolor=RoyalBlue,urlcolor=RoyalBlue}

\usepackage{comment}
\usepackage{appendix}
\usepackage{graphicx}

\usepackage{soul}
\usepackage{tikz}
\usepackage{pgfplots}
\usepackage{dcolumn}
\usepackage{bm}
\usepackage{amsmath}

\usepackage[mathlines]{lineno}
\usepackage{xcolor}
\usepackage{slashed}

\newcommand{\spindersphere}{{}_s\slashed{\Delta}}
\usepackage[normalem]{ulem}

\pgfplotsset{compat=1.17}
\begin{document}

\title{Challenges in Quasinormal Mode Extraction: Perspectives from Numerical solutions to the Teukolsky Equation}

\author{Hengrui Zhu\orcidlink{0000-0001-9027-4184}}
\email{hengrui.zhu@princeton.edu}
\affiliation{Department of Physics, Princeton University, Jadwin Hall, Washington Road, New Jersey, 08544, USA}
\affiliation{Princeton Gravity Initiative, Princeton University, Princeton, New Jersey, 08544, USA}

\author{Justin L. Ripley\orcidlink{0000-0001-7192-0021}}

\affiliation{Illinois Center for Advanced Studies of the Universe \& Department of Physics, University of Illinois at Urbana-Champaign, Urbana, IL 61801, USA}

\author{Alejandro Cárdenas-Avendaño\orcidlink{0000-0001-9528-1826}}
\affiliation{Department of Physics, Princeton University, Jadwin Hall, Washington Road, New Jersey, 08544, USA}
\affiliation{Princeton Gravity Initiative, Princeton University, Princeton, New Jersey, 08544, USA}
 
\author{Frans Pretorius}
\affiliation{Department of Physics, Princeton University, Jadwin Hall, Washington Road, New Jersey, 08544, USA}
\affiliation{Princeton Gravity Initiative, Princeton University, Princeton, New Jersey, 08544, USA}
\date{\today}

\begin{abstract}

The intricacies of black hole ringdown analysis are amplified by the absence of a complete set of orthogonal basis functions for quasinormal modes. Although damped sinusoids effectively fit the ringdown signals from binary black hole mergers, the risk of overfitting remains, due to initial transients and nonlinear effects. In light of this challenge, we introduce two methods for extracting quasinormal modes in numerical simulations and qualitatively study how the transient might affect quasinormal mode fitting. In one method, we accurately fit quasinormal modes by using their spatial functional form at constant time hypersurfaces, while in the other method, we exploit both spatial and temporal aspects of the quasinormal modes. Both fitting methods leverage the spatial behavior of quasinormal eigenfunctions to enhance accuracy, outperforming conventional time-only fitting techniques at null infinity. We also show that we can construct an inner product for which the quasinormal eigenfunctions form an orthonormal (but not complete) set. We then conduct numerical experiments involving linearly perturbed Kerr black holes in horizon penetrating, hyperboloidally compactified coordinates, as this setup enables a more precise isolation and examination of the ringdown phenomenon. 
From solutions to the Teukolsky equation, describing scattering of an ingoing gravitational wave pulse,
we find that the contributions from early-time transients can lead to large uncertainties in the fit to the amplitudes of higher overtones ($n\geq 3$).
While the methods we discuss here cannot be applied directly to data from merger observations, our findings underscore the persistence of ambiguities in interpreting ringdown signals, even with access to both temporal and spatial information.
\end{abstract}

\maketitle

\section{\label{sec:level1}Introduction}

According to general relativity, when two black holes merge they form a highly distorted black hole that then rings down a stationary Kerr black hole. 
The gravitational waves emitted during the ringdown are thought to be well described by linear black hole perturbation theory (\emph{linear theory} for short), as governed by the spin-$\pm2$ Teukolsky equation~\cite{Teukolsky:1973ha,Teukolsky:1974yv}.
The Teukolsky equation is a separable partial differential equation for a single complex scalar, the real and imaginary parts of which describe the two physical gravitational wave polarizations. 

The Teukolsky equation (with physical boundary conditions) does not have mode solutions; instead it has exponentially decaying {\em quasinormal mode} (QNM) solutions \cite{Kokkotas:1999bd,Berti:2009kk}. 
It is thought that shortly after merger the ringdown signal is well described by a superposition of QNMs~\cite{Vishveshwara:1970zz,Press:1971wr,Price:1994pm}, and fits to numerical relativity results seem to confirm this expectation (e.g.,~\cite{Buonanno:2006ui, Berti:2007fi, LIGOScientific:2016vlm,Giesler:2019uxc,Li:2021wgz}). 
At much later times ($\mathcal{O}(100M)$ after coalescence, where $M$ is the mass of the remnant), the signal is expected to transition to decay as a power law. 
This is the so-called ``tail'' part of the waveform~\cite{PhysRevD.5.2419,Krivan:1997hc,Scheel:2003vs,Tiglio:2007jp}. 
An interesting property of the frequencies of the QNMs is that they are uniquely determined by the remnant black hole's mass and spin. 
The reverse process of estimating the mass and spin of the remnant black hole from the QNM spectrum of the ringdown is known as \emph{black hole spectroscopy}~\cite{Detweiler:1980gk,Dreyer:2003bv,Berti:2005ys,Isi:2020tac, Carullo:2019flw, Isi:2021iql}.

To maximize the science that can be extracted from an observed ringdown---whether for measuring properties of the merger, or for testing general relativity---one needs a prediction for what the excitation amplitude of each QNM is for a given merger. 
At present, computing these excitation amplitudes is an open problem for a remnant formed in a merger of black holes with comparable mass (though some information can be gleaned from properties of the Green's function for Kerr perturbations~\cite{berti:2006kk}, or using linear theory in the extreme mass ratio limit~\cite{Hughes:2019zmt,Apte:2019txp,Lim:2019xrb}). 
In lieu of such calculations, one can attempt to measure the excitation amplitudes directly from numerical relativity solutions of merger events. 
At present, such approaches typically assume the ringdown can be entirely described by a sum of linear QNMs, and attempt to find the best first set of amplitudes that reproduce the ringdown signal (see e.g.,~\cite{Berti:2007fi,Buonanno:2006ui,Giesler:2019uxc,Siegel:2023lxl, Zhu:2023fnf,Kamaretsos:2012bs}). 
These studies have demonstrated, for example, that in an astrophysical merger of a nearly equal mass, non-precessing binary, the $l=m=2$ mode is the maximally excited QNM, and the relative excitation amplitudes of other angular modes may point to properties of the progenitor binary system, e.g., precession, eccentricity, mass ratio, etc. 

There are many difficulties in attempting to ascribe excitation amplitudes to merger events from fits to numerical relativity waveforms. 
The main difficulty already present at the linear level is that the QNMs do not comprise a complete basis of black hole perturbations, and the gravitational wave ``perturbation'' will contain a {\em transient part} that can have a significant amplitude relative to the sum of QNMs, in particular in the time window of $O(10M)$ around peak amplitude of the ringdown. 
Note that in this paper we use the phrase {\em transient part} (or sometimes {\em prompt part}) to refer to the non-QNM part of a ringdown waveform~\cite{Baibhav:2023clw}. 
Beyond the linear level a host of additional difficulties arise, including non-linear mode coupling (quadratic modes have only recently begun to be studied in full numerical relativity merger waveforms~\cite{London:2014cma,Ma:2022wpv,Mitman:2022qdl,Cheung:2022rbm,Khera:2023lnc}), and the effects of back reaction of the gravitational wave energy~\cite{Sberna:2021eui}. 
The latter complicates the questions of what the appropriate background is for computing linear perturbations about, and how good a constant amplitude approximation is for the early time linear QNM spectrum of the waveform (due, in part, to non-linear energy exchange between modes). 
Though these difficulties are not thought to have much effect on measuring the dominant fundamental $l=m=2$ QNM, it is less clear how well higher overtones and harmonics can be extracted. 
As such there is still much debate within the gravitational wave community about which modes should be included in the ringdown fit (see e.g.,~\cite{Giesler:2019uxc,Baibhav:2023clw}). 

Given the intrinsic complexity of the problem and since both non-modal and nonlinear effects could play a nontrivial role, several ways of analyzing and decomposing the ringdown signal from numerical simulations into QNMs have been proposed~\cite{Giesler:2019uxc,Li:2021wgz,Ma:2022wpv,London:2022urb,Baibhav:2023clw}. 
Most of these methods involve finding the best fit to the ringdown signal with a sum of damped sinusoids with quasinormal mode frequencies\footnote{An exception to this procedure is Ref.~\cite{Ma:2022wpv}, where the authors eliminated the dominant modes through a linear filter. Another exception is Ref.~\cite{London:2022urb}, where properties of spheroidal harmonics are explored to separate the prograde and retrograde contribution to the ringdown signal.}, using gravitational waveforms extrapolated to future null infinity, or through Cauchy-characteristic extraction (CCE). Though, as discussed above, the signal is expected to contain more than simply the set of linearly damped QNMs, and if we do not know {\em a priori} what the transient part of the waveform is, it is easy to envision that this process could result in {\em overfitting} : an erroneous QNM amplitude measurement due to overlap of the QNM mode with a transient part of the signal. Particularly susceptible to overfitting are the higher overtones, whose damping times are sufficiently rapid that unless they are excited with extremely high amplitude, only a small number of cycles, or even only a  fraction of a cycle of the QNM will be visible above an effective noise floor set by numerical truncation error (assuming all other extraction systematics are under control). Some studies have already pointed to overfitting, by showing different fitting procedures can give different results for the QNM amplitudes of the ringdown of a black hole produced from a binary collision~\cite{Baibhav:2023clw,PhysRevD.108.044032}.

The main purpose of this paper is to gain more insight into the nature of mode fitting, and hence the problem of overfitting.
Instead of studying the full nonlinear ringdown of a black hole produced from a binary collision, we attempt to reconstruct the quasinormal mode spectrum of solutions to the Teukolsky equation. 
This allows us to study in detail how easy it is to distinguish the transient contribution to the signal from the quasinormal modes\footnote{We expect the transient contribution to strongly depend on the initial data; here, we only focus on scattering experiment, where the initial data consist of an infalling pulse of gravitational wave onto the black hole.}. 
In our fitting procedures, we include the spatial profiles of the quasinormal mode eigenfunctions, which reduce systematic uncertainties in our fits. 
To aid in utilizing the spatial dependence of the QNMs in our fits we make use of horizon penetrating, hyperboloidally compactified (HPHC) coordinates, in which the QNM solutions to the Teukolsky equation are regular everywhere, from the black hole horizon to null infinity \cite{zenginouglu2011geometric,Ripley:2022ypi}.
We consider two fitting procedures to linear data: one that uses the spatial variation of the Weyl scalar field and its time derivative on a single constant time slice, and another that uses both spatial and temporal information. 
Within both procedures, fitting the quasinormal mode amplitudes reduces to a problem of linear regression, \emph{given} (as we have here) exact knowledge of the background black hole mass and spin. 

We then apply these fitting procedures to a set of time domain numerical solutions to the Teukolsky equation.
We demonstrate that with pure QNM initial data, we can stably recover the amplitudes of arbitrary linear combinations of QNMs.
By {\em stable} here we mean that the method recovers the correct amplitudes (to within truncation error) over a non-negligible window of time. 
When we consider scattering initial data with non-QNM contributions though, we find that we cannot stably extract the amplitude of higher ($n\geq 3$) QNM overtones, and the traditional time-only fit at future null infinity can only faithfully stably extract the fundamental and first overtone over a much narrower window for fitting time. 
Conversely, we demonstrate the power of using spatial information to establish a best-case scenario for extracting QNMs.
We note that this paper is more a ``proof of principle'' for the linear case, in that we have not tried to optimize the linear perturbation to ``best fit'' any particular merger event. We leave a more extensive study of the issue of initial conditions to future work.

The rest of this paper is organized as follows. In Sec.~\eqref{sec:teuk}, we review the derivation of the Teukolsky equation in HPHC coordinates, our code for computing pure QNM initial data, and our code for evolving the Teukolsky equation in the time domain.
In Sec.~\eqref{sec:fit}, we introduce our two fitting procedures that make use of spatial information of the quasinormal modes. 
In Sec.~\eqref{sec:numerical-results}, we show results from applying those two methods to numerical solutions to the Teukolsky equation with several different classes of initial data.
Finally, we compare our new fitting procedures with the traditional time-only fit at future null infinity in Sec.~\eqref{sec:time-only}.
We discuss the implications of our results and conclude in Sec.~\eqref{sec:conclude}. In Appendices \ref{app:qne}, \ref{app:qnm_radial}, and \ref{app:convergence} we discuss some details of computing the QNM eigenfunctions, their radial structure in HPHC coordinates, and give some convergence results from our code, respectively.


\section{The Teukolsky equation on a hyperboloidal slice}
\label{sec:teuk}

In this section, we briefly review the Teukolsky equation and QNMs in HPHC coordinates. 
We refer the reader to  Refs.~\cite{zenginouglu2011geometric, Harms:2013ib, PanossoMacedo:2019npm, Ripley:2020xby, Ripley:2022ypi} for further details. 

The Teukolsky equation (TE)~\cite{Teukolsky:1973ha} was first derived in Boyer-Lindquist (BL) coordinates~\cite{doi:10.1063/1.1705193}. 
Constant time hypersurfaces in BL coordinates do not penetrate the black hole horizon, nor do they reach future null infinity -- instead, they stretch from the bifurcation sphere of the black hole to spatial infinity. 
One consequence of these properties is that the radial eigenfunctions for quasinormal modes are necessarily divergent at the asymptotic radial boundaries ($r_*\to\pm\infty$, where $r_*$ is the tortoise coordinate) when evaluated on constant time slices \cite{Teukolsky:1973ha}.
This feature of the quasinormal eigenfunctions (QNEs) in BL coordinates complicates the analysis of computing QNM excitation factors of black hole ringdown.
This is because constructing a well-defined inner product (from which the excitation factors of the quasinormal modes can be computed) involves an integration in the complex plane \cite{PhysRevD.34.384,PhysRevD.38.1040,berti:2006kk,Green:2022htq,Ansorg:2016ztf}. 
By contrast, since constant time hypersurfaces in HPHC coordinates span from the black hole horizon to future null infinity, the QNM solutions to the TE in these coordinates remain regular everywhere exterior to the black hole \cite{zenginouglu2011geometric,Ripley:2022ypi}.
This opens up the possibility of a simpler inner-product matrix that could be used to determine the quasinormal mode content of a given gravitational waveform (see, for example, Ref.~\cite{Jaramillo:2020tuu}).
Furthermore, the ringdown signal behaves like damped standing waves spatially in HPHC, instead of traveling wave packets in coordinates that asymptote to spatial infinity. 

In this work we use the same HPHC coordinates described in Ref.~\cite{Ripley:2022ypi}. 
These coordinates are identical to BL coordinates, up to a redefinition of the time coordinate $\tau$ and azimuthal coordinate $\phi$, which are related to the BL coordinates $(t,r,\vartheta,\varphi)$ via
\begin{eqnarray}
   d\tau
   \equiv
   dt
   +
   \left(\frac{2Mr}{\Delta} + \frac{dh}{dr}\right)dr
   ,\qquad
   d\phi
   \equiv
   d\varphi
   +
   \frac{a}{\Delta}dr
   ,
\end{eqnarray}
where $M$ and $a$ are the mass and spin of the black hole.
Here $h(r)$ is a ``height'' function designed to make the radially ingoing characteristic speed zero at infinity \cite{zenginouglu2008hyperboloidal,PanossoMacedo:2019npm,Ripley:2020xby}, which we chose to be
\begin{equation}
   \frac{dh}{dr} = - 1 - \frac{4M}{r}.
\end{equation}
To bring future null infinity (located at $r\to\infty)$ to a finite point, we compactify the radial coordinate via
\begin{equation}\label{eq:radial}
   \rho \equiv \frac{1}{r}.
\end{equation}
We additionally rescale the Newman-Penrose scalar $\psi$ to make the Teukolsky equation regular at the horizon and to remove the ``long-range potential'' in the radial coordinate~\cite{SASAKI198268,Hughes:1999bq}
\begin{equation}\label{eq:Psi4}
   \psi
   \equiv
   \frac{1}{r}\Delta^{-s} \Psi
   .
\end{equation}
With all the above definitions, the TE reads
\begin{widetext}
\begin{eqnarray}
\label{eq:hyperboloidal_teuk_eqn}
   \left[
      16M^2 -
      a^2\sin^2\theta
      +
      8M\left(4M^2-a^2\right)\rho
      -
      16a^2M^2\rho^2
   \right]
   \partial_{\tau}^2\Psi
   -
   \rho^4\Delta\partial_{\rho}^2\Psi
   -
   \spindersphere\Psi
   \nonumber
   \\
   -
   2\left[
      1
      +
      \left(a^2 - 8M^2\right)\rho^2
      +
      4a^2M\rho^3
   \right]
   \partial_{\tau}\partial_{\rho}\Psi
   +
   2a\rho^2\partial_{\rho}\partial_{\phi}\Psi
   +
   2a\left(1+4M\rho\right)\partial_{\tau}\partial_{\phi}\Psi
   \nonumber
   \\
   +
   2
   \left[
      s\left(
         -
         2M
         +
         ia\cos\theta
      \right)
      +
      \left(
         4M^2\left\{s+2\right\}
         -
         a^2
      \right)
      \rho
      -
      6Ma^2\rho^2
   \right]
   \partial_{\tau}\Psi
   \nonumber
   \\
   +
   2
   \left[
      -
      1
      -
      s
      +
      \left(s+3\right)M\rho
      -
      2a^2\rho^2
   \right]
   \rho
   \partial_{\rho}\Psi
   +
   2a\rho\partial_{\phi}\Psi
   \nonumber
   \\
   +
   2\left(
      Ms
      +
      M
      -
      a^2\rho
   \right)
   \rho
   \Psi
   &=
   0,
\end{eqnarray}
\end{widetext}
where $s$ is the spin-weight of the scalar $\Psi$.
For the remainder of this article, we set $s=-2$, so that $\Psi$ corresponds to the Weyl scalar $\Psi_4$.

Finally, to make the radial boundary independent of the black hole spin, we perform the substitution:
\begin{equation}
    \rho\to  r_+\rho,
\end{equation}
where $r_+=M+\sqrt{M^2-a^2}$ is the radius of the outer horizon in BL coordinates. 
This substitution makes the TE regular at future null infinity ($\rho=0$) and on the black hole horizon ($\rho=1$), regardless of spin. 

We solve Eq.~\eqref{eq:hyperboloidal_teuk_eqn} in the time domain, using a modification of the code described in Refs.~\cite{Ripley:2020xby,justin_ripley_2023_8215611}, which we will now briefly describe. 
The numerical implementation decomposes $\Psi$ into its azimuthal modes, $\Psi\left(t,\rho,\theta,\phi\right) = \sum_m e^{im\phi}\Psi\left(t,\rho,\theta\right)$.
The code then evolves each $m-$mode on a two dimensional $\rho-\theta$ grid. 
The angular direction is discretized using a pseudospectral method made up of spin-weighted spherical harmonics, and the radial direction with a fourth order finite difference method, as opposed to the implementation presented in Ref.~\cite{Ripley:2020xby}, which makes use of a pseudospectral Chebyshev discretization in the radial direction. 
To evolve in time, the code uses a method-of-lines algorithm with a fourth-order Runge Kutta integrator. 
We consider two classes of initial data, described in more detail in Sec.~\eqref{sec:numerical-results}: (1) a linear superposition of quasinormal modes, and (2) a Gaussian pulse (which we call ``scattering'' initial data).
We construct our quasinormal mode initial data using a slight modification, described in detail in Appendix~\eqref{app:qne}, of the algorithm presented in Ref.~\cite{Ripley:2022ypi} (publicly available at~\cite{justin_ripley_2023_8215611}.)

\section{Spatial and spacetime fitting with QNM eigenfunctions} \label{sec:fit}

Let us consider a linearly perturbed black hole with fixed known mass and spin. 
Since the quasinormal mode decomposition of the solution can be recovered using a linear least-squares algorithm if the linearized gravitational solution can be entirely described as a superposition of quasinormal modes~\cite{Giesler:2019uxc,Mitman:2022qdl}, we fix the quasinormal mode frequencies, and then fit for the complex amplitudes of the modes that minimize the residual error. 
In our fitting procedures, we minimize not just the residual error of our waveform fit at future null infinity, but also the error of the waveform over the entire computational domain, which ranges from the horizon to null infinity. 

We consider two different mode extraction methods: \emph{spatial} and \emph{spacetime} fitting, which we describe in detail in Secs.~\eqref{sec:exc} and~\eqref{eq:spacetime-fit}, respectively. 
Spatial fitting refers to measuring the amplitudes for each QNM on a fixed time slice $t=t_0$, given the data $\{\Psi_4(t_0,\mathbf{r}),\partial_t \Psi_4(t_0,\mathbf{r})\}$~\cite{berti:2006kk,Cheung:2022rbm,Zhang:2013ksa}. 
That is, for a fixed azimuthal number $m$, we minimize the residual 
\begin{widetext}
\begin{align}\label{eq:res_ext}
    \mathcal{R}
    =&
    \sum_{i,j}\left(
        \Psi_4\left(t_0,\rho_i,\theta_j\right)
        -
        \sum_{[p],n,l} A_{[p]ln} R_{[p]ln}\left(\rho_i\right){}_{-2}S\left(a\omega_{[p]ln},\theta_j\right) e^{-i\omega_{[p]ln}t_0}
    \right)^2 \nonumber \\
    &+
    \left(
        \partial_t \Psi_4\left(t_0,\rho_i,\theta_j\right)
        +
        \sum_{[p],n,l} i\omega_{[p]ln}A_{[p]ln} R_{[p]ln}\left(\rho_i\right){}_{-2}S\left(a\omega_{[p]ln},\theta_j\right) e^{-i\omega_{[p]ln}t_0}
    \right)^2 
    ,
\end{align}
\end{widetext}
for the complex constants $A_{[p]ln}$, where $_{-2}S$ are the spin-weighted \emph{spheroidal} harmonics, and $R_{[p]ln}(\rho)$ and $\omega_{[p]l'n}$ are the QNM  radial eigenfunctions and frequencies, respectively. 
In the above expression, the sum is over the prograde and retrograde ($[p]=\pm$) modes, the overtones $n$, angular number $l$, radial grid points $\rho_i$, and angular grid points $\theta_j$. 
In practice, we perform a spherical harmonic decomposition of the signal in $\theta$ before minimizing the residual.

On the other hand, the spacetime fitting consists of finding the best quasinormal mode fit to the rescaled Weyl scalar $\Psi_4$ over the entire time domain we evolve for, i.e., in \emph{both} space and time. Specifically, we minimize the residual
\begin{widetext}
\begin{align}\label{eq:res_fit}
    \mathcal{R}
    &=
    \sum_{i,j,k}\left(
        \Psi_4\left(t_k,\rho_i,\theta_j\right)
        -
        \sum_{[p],n,l} A_{[p]ln} R_{[p]ln}\left(\rho_i\right){}_{-2}S\left(a\omega_{[p]ln},\theta_j\right) e^{-i\omega_{[p]ln}t_k}
    \right)^2~,
\end{align}
\end{widetext}
where now we include a sum over the time steps $t_k$. As we discussed above, both fitting methods differ from previous QNM fitting procedures as our residual includes the radial profile of the modes. 

If the gravitational waveform is dominated by quasinormal modes, our fitting procedure provides a robust way to determine the quasinormal mode content of a gravitational waveform. 
We now provide specific details of both approaches. 

\subsection{Spatial Fitting} \label{sec:exc}

In this approach, we find a sum of the QNE with amplitudes that best represent the data $\{\Psi,\partial_t \Psi\}$ on a constant time hypersurface~\cite{berti:2006kk,Berti:2006hb,Dorband:2006gg}. 
At intermediate times $t$, i.e. after initial data transients have decayed but before the tail contributions are evident, we expect the linear gravitational wave to be well approximated by a sum of quasinormal modes. 
In this regime, the field and its time derivative on a constant time slice, $t_0$, can then be approximated by:
\begin{widetext}
\begin{equation}\label{eq:mode}
    \Psi_4(\rho,\theta,t_0) = \sum_{p\in\{\pm\}}\sum_n\sum_{l}A_{[p]ln} ~_{-2}Y_{l}(\theta)\sum_{l'}c_{[p]ll'n}~ R_{[p]l'n}(\rho)\exp\{-i\omega_{[p]l'n} t_0\}~,
\end{equation}
\begin{equation}\label{eq:dtmode}
    \partial_t\Psi_4(\rho,\theta,t_0) = \sum_{p\in\{\pm\}}\sum_n\sum_{l}A_{[p]ln} ~_{-2}Y_{l}(\theta)\sum_{l'}\left(-i\omega_{[p]l'n}\right)c_{[p]ll'n}~ R_{[p]l'n}(\rho)\exp\{-i\omega_{[p]l'n} t_0\}~,
\end{equation}
\end{widetext}
where 
$c_{[p]ll'n}$ are the spherical-spheroidal mixing coefficients, $_{-2}Y_{l}$ are the spin-weighted \emph{spherical} harmonics, and $R_{[p]ln}(\rho)$ and $\omega_{[p]l'n}$ are the QNM eigenfunctions and frequencies, respectively. 

We can rewrite Eqs.~\eqref{eq:mode} and~\eqref{eq:dtmode} as a matrix equation for the amplitudes $A_{[p]ln}$. 
In terms of the spherical harmonics for $\Psi_4$, we may write for each angular number $l$, 
\begin{subequations}
\label{eq:id-problem}
\begin{align}
    M_{[p]ll'n}(\rho_i)A_{[p]l'n} &= \Psi_{4,l}(\rho_i) \label{eq:matrix} \\ 
    -i\omega_{[p]l'n}M_{[p]ll'n}(\rho_i)A_{[p]l'n} &= \partial_t\Psi_{4,l}(\rho_i)~, \label{eq:dtmatrix}
\end{align}
\end{subequations}
where repeated indices are summed over\footnote{The $i$'s in parenthesis and as subscripts index the radial grid points, $\sqrt{-1}$ otherwise. }, and 
\begin{align}
    \Psi_{4,l}(\rho,t) 
    &:=
    \int_\theta \Psi_4(\rho,\theta,t) ~_{-2}Y_l^*(\theta) d \theta
    ,\\
    M_{[p]ll'n}(\rho_i) 
    &:= c_{[p]ll'n}~ R_{[p]l'n}(\rho_i)\exp\{-i\omega_{[p]l'n} t_0\}~.
\end{align}

The QNM amplitudes $A_{[p]l'n}$ must simultaneously solve Eqs.~\eqref{eq:matrix} and \eqref{eq:dtmatrix} for all $l$, which we do numerically. 
Here, we can simply stack the two matrix equations in the radial direction (indexed by $i$) and solve the resultant equation by a minimization matrix solver. 
Specifically, we stack via the following rule 
\begin{equation}\label{eq:excitation_design}
    N_{[p]ll'n}(i)=
    \begin{cases}
        M_{[p]ll'n}(\rho_i) & \text{if } i \leq i_{max}\\
        -i\omega_{[p]l'n} M_{[p]ll'n}(\rho_{i-i_{max}}) & \text{if } i>i_{max}, \\
    \end{cases}
\end{equation}
and, similarly, for the right-hand side of Eqs.~\eqref{eq:id-problem}:
\begin{equation}\label{eq:rhs}
    b_l(i)=
    \begin{cases}
        \Psi_{4,l}(\rho_i) & \text{if } i \leq i_{max}\\
        \partial_t\Psi_{4,l}(\rho_{i-i_{max}}) & \text{if } i>i_{max}~,\\
    \end{cases}
\end{equation}
where $i_{max}$ is the number of radial grid points. Under this procedure, Eqs.~\eqref{eq:id-problem} can now be written as
\begin{align}
\label{eq:matrix-eqn}
    N_{IJ}A_{J} = b_I,
\end{align}
where $I$ indexes the spatial components (radial $i$ and angular $l$,) and $J$ indexes the modes (prograde/retrograde $[p]$, angular index $l^{\prime}$, and overtone number $n$.) 
The matrix $N_{IJ}$, where we pack the fitting basis functions as column vectors, is called the \emph{design matrix} (see, e.g., Ref.~\cite{2004Geode.121..155M}).

We find that, when the initial data are a pure, arbitrary superpositions of QNMs, we correctly recover the amplitudes and phases of the modes when we solve the matrix equation \eqref{eq:matrix-eqn}. 
The design matrix induces an inner product via
\begin{align}
\label{eq:bilinear-form}
P(\Psi_{[p]ln},\Psi_{[p']l'n'}) &:= \frac{1}{2} \langle N^{-1}\Psi_{[p]ln},N^{-1}\Psi_{[p']l'n'}\rangle    \nonumber
\\ &=\delta_{pp'}\delta_{ll'}\delta_{nn'},
\end{align}
 where $\langle\cdot,\cdot\rangle$ denotes the usual inner product on $\mathbb{C}^d$, and $\Psi_{[p]ln}$ and $\Psi_{[p']l'n'}$ are constructed from the quasinormal mode eigenfunctions, as in Eq.~\eqref{eq:rhs}.
We numerically find that the design matrix defined in Eq.~\eqref{eq:excitation_design} has full rank and its right inverse exists as long as the overtones are radially resolved, with $i_{max} \gg n_{max}$, i.e., there are more radial points than overtones in our fit. 
For given numerical data, we can determine the QNM amplitudes by computing $N^{-1}b$. 

\subsection{Spacetime Fitting}
\label{eq:spacetime-fit}

We now minimize a quadratic residual, as with the spatial fitting, but now we also sum over the different time steps. 
When the linear gravitational wave is dominated by QNMs, the fitting problem reduces again to solving Eq.~\eqref{eq:mode} for the amplitude $A_{[p]ln}$'s, given numerical data $\left\{\Psi_4(\rho, \theta ,t)\right\}$ within $t\in[t_0,t_1]$. 

As in the spatial fit, we decompose $\Psi_4$ into spin-weighted spherical harmonics. 
Discretizing the radial ($r_i$) and time ($t_j$) coordinates, the design matrix now takes the form:
\begin{align}
    &M_{[p]ll'n} (\rho_i,t_j) \nonumber = \\
    & ~\left(-i\omega_{[p]l'n}\right)c_{[p]ll'n}~ R_{[p]l'n}(\rho_i)\exp\{-i\omega_{[p]l'n} t_j\}~,
\end{align}
where now the right-hand side is set to be the field for the entire spacetime perturbation:
\begin{equation}
    b_l(\rho_i,t_j) = \Psi_{4,l}(\rho_i,t_j)~.
\end{equation}
The spacetime fitting as a matrix equation is then 
\begin{align}
\label{eq:matrix-eqn-fit}
    M_{IJ}A_{J} = b_I,
\end{align}
where $I$ now indexes \emph{both} the temporal and spatial components (time $j$, radial $i$, and angular $l$), and $J$ indexes the modes by $[p]$, $l^{\prime}$, and $n$. 
In Secs.~\eqref{sec:multi_qnm} and~\eqref{sec:scatter} we demonstrate that the spacetime fit results are consistent with the spatial fit in regimes where we expect QNMs to dominate the solution.

Before going to the numerical examples, we first briefly discuss the incompleteness of the QNEs as a function basis for general solutions to the TE, and a resulting caveat of fitting QNMs due to the presence of non-modal solutions of the TE. 

\subsection{A Caveat: Mode vs Transient}
\label{sec:caveats-modes-transiets}

While we can define an inner product under which the QNMs are orthonormal by making use of their radial and angular information, the modes remain incomplete as a basis for fitting black hole ringdown.
By incompleteness, we mean that a generic solution to the TE cannot be represented as a sum of QNMs, when the solution violates the physical boundary condition: no ingoing wave at the horizon and no ingoing wave at infinity\footnote{Reference~\cite{London:2023aeo} suggests completeness of QNMs as a basis for solutions to the TE that respect the physical boundary conditions.}. 
As we already mentioned, in addition to QNMs, solutions to the TE also admit a ``prompt'' and a ``tail'' contribution \cite{Leaver:1985ax,Nollert:1999ji}, the sum of which we refer to as the ``transient'' part of the solution.
Prompt here relates to the kind of perturbations we expect following a black hole merger, or scattering a compact wave packet off the black hole, and refers to the early rise in the waveform before the QNMs dominate.
The tail part of the solution arises from backscattered gravitational waves on the Kerr geometry, and dominates the solution at late times (beyond the times considered in this paper).

At the linear level there are no prompt or tail contributions to the solution if the initial data consists of purely quasinormal modes.
However, for more generic initial data that better describes a distorted black hole formed from an astrophysical merger, there will be prompt and tail contributions~\cite{berti:2006kk,Baibhav:2023clw}. 
In these more generic settings, assuming the signal is purely made up of QNMs, fitting to those can lead to biased results, in particular, for the high-$n$ overtones, as they typically decay quite rapidly and on similar timescales to the prompt part of the transients. 
As we mentioned earlier, we call this overfitting the signal. 

The prompt response dies off rapidly in time as it is sourced over a relatively small spacetime volume around the remnant black hole at the time of merger, and the corresponding wave fronts essentially follow geodesic trajectories to either future null infinity or the black hole horizon. Starting the QNM fit at later times should reduce the bias caused by the contribution of this transient response in the signal. However, the exact form of prompt response depends heavily on the initial data; in some cases one might expect it to be large enough, and decay slowly enough, to mask the higher overtones.
By contrast, the tail contribution decays in a power-law fashion in time, slower than the QNM contribution~\cite{PhysRevD.5.2419}. Thus, the tail response may bias quasinormal mode fitting at late times (provided the signal-to-noise ratio of the signal was large enough to resolve a late-time signal). 

To assess the quality of our fitting results when non-QNM contributions to the solution are present, we adapt the technique presented in Refs.~\cite{Baibhav:2023clw,Mitman:2022qdl,Cheung:2022rbm,Bhagwat:2019dtm,PhysRevD.108.044032}. 
Namely, we vary the start time of the spacetime fitting, or time at which we apply the spatial fitting, and check if the amplitude for each quasinormal mode remains constant. 
We discuss the results of this exercise in Secs.~\eqref{sec:multi_qnm} and~\eqref{sec:scatter}. 

\section{Numerical examples} \label{sec:numerical-results}

Here we present some examples of applying the proposed spatial and spacetime fitting to numerical solutions of the TE with different initial conditions, as described in Sec.~\eqref{sec:teuk}. 
Unless otherwise mentioned, all simulation results presented here were from runs with resolution $n_\rho=512$ and $n_\theta=40$, where $n_\rho$ and $n_\theta$ are the number of radial and angular grid points, respectively (see Appendix~\ref{app:convergence} for convergence results).

First, in Sec.~\eqref{eq:pure-qnm} we evolve initial data that consist of a single QNM, to demonstrate the accuracy of our evolution and (quasinormal mode) initial data codes, described in the Appendix~\eqref{app:qne}. 
In Sec.~\eqref{sec:multi_qnm}, we move to a more complicated class of initial data: a superposition of QNMs. 
In this case, we demonstrate that we can still reliably recover the amplitudes of the QNMs up to numerical precision of the solution, using both fitting techniques. 

Finally, in Sec.~\eqref{sec:scatter} we consider scattering initial data (that is, initial data that cannot be written as a pure sum of quasinormal modes). 
In this case, we also extract the QNM content from the signal, although we do not have a direct theoretical estimate for the QNM amplitudes for this class of initial data. 
We do demonstrate that both the spatial and spacetime fitting methods are consistent, in the sense that both yield identical estimates for the QNM amplitudes given the same initial data, and such estimates are stable with respect to fitting time at least for the fundamental mode and the first two overtones\footnote{Due to how long-lived the transient is in the case of a near-extremal black hole, for that example we can only extract the fundamental mode during the numerical integration time of $500$M. }. 
We further point out that the instability of fitting to the $n\geq 3$ overtones for scattering initial data is likely due to the presence of the transient solution masking the high-$n$ overtone spectrum, though their initial excitation amplitudes might be lower than that from black hole mergers. 

\subsection{Evolving a single QNM\label{eq:pure-qnm}}

\begin{figure*}[t]%
  \includegraphics[width=\textwidth]{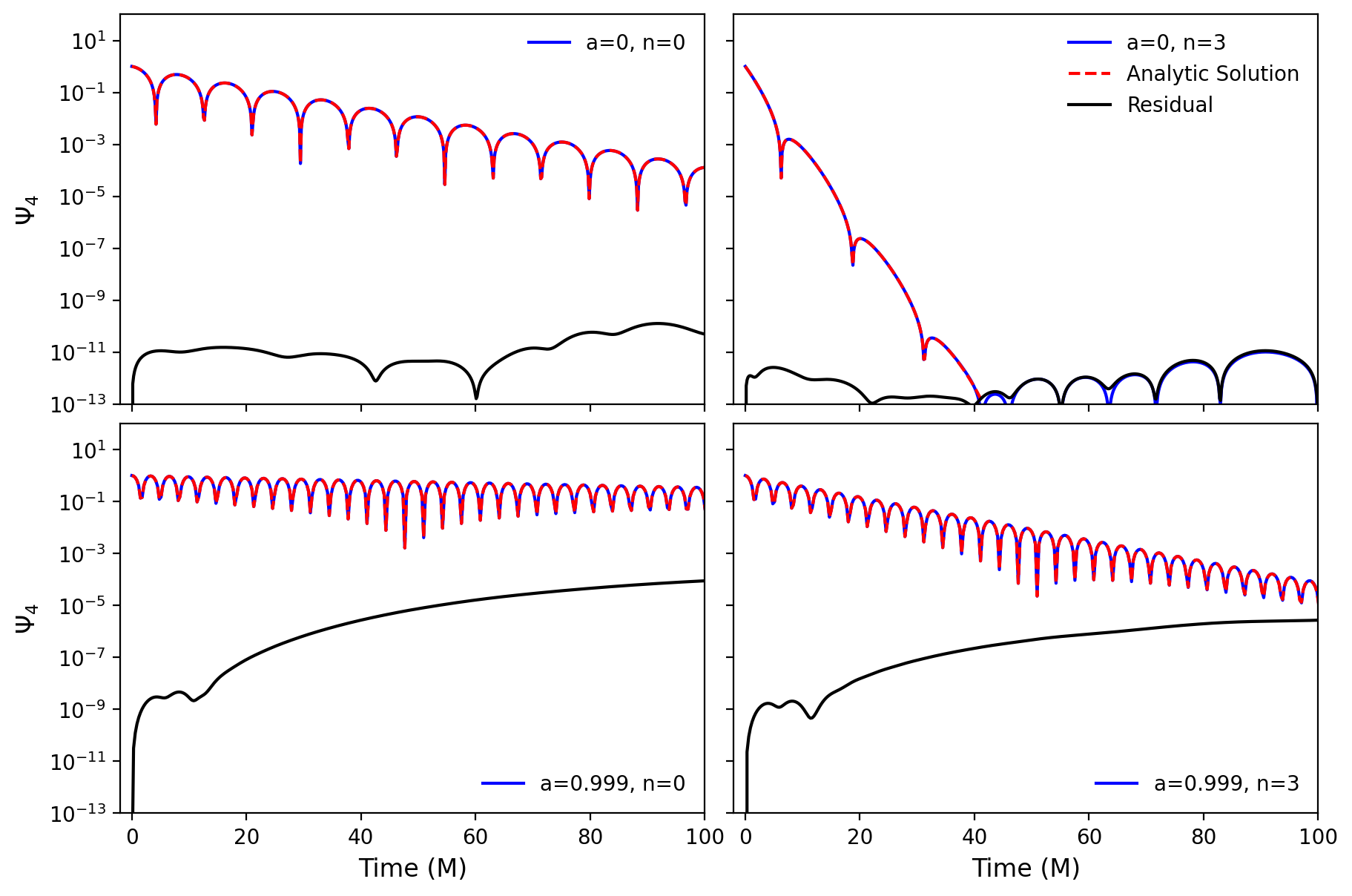}
  \caption{Evolving single QNMs ($n=0$ and $n=3$, left and right panels, respectively) for (top) Schwarzschild  and (bottom) near-extremal Kerr with $a=0.999$. The numerical solution at future null infinity is shown with solid blue lines, while the analytical predictions are drawn with red dashed lines. 
  Plotted in black is the residual/difference between the two, which convergence studies (see Appendix~\eqref{app:convergence}) show arises purely from numerical truncation error.}
  \label{fig:pure_qnm}
\end{figure*}

Let us consider the evolution of a single QNM for both Schwarzschild and near-extremal Kerr ($a=0.999$) backgrounds. 
We set the initial condition to be either the $l=m=2$ fundamental mode ($n=0$), or the $l=m=2,~n=3$ overtone\footnote{With our implementation we can evolve even higher overtones accurately; we only show the $n=3$ overtone as an example.}. 
As illustrated in Fig. ~\eqref{fig:pure_qnm}, we can accurately evolve both the fundamental mode and overtone (blue solid lines) as compared to the analytic solution (red dashed lines). 
The residuals at future null infinity between the analytical solution and these runs are plotted in black. 
As shown in Appendix~\eqref{app:convergence}, this residual converges to zero at fourth order with increasing resolution.

These results are a strong test of our evolution code: if an overtone is excited, the code is capable of capturing it up to numerical precision. 
This accuracy provides the foundation for our following analysis. 

\subsection{Evolving and fitting to a superposition of QNMs} \label{sec:multi_qnm}

In this section, we consider initial data that consist of a superposition of QNMs. 
We demonstrate that the spatial and spacetime fitting procedures, proposed in Sec.~\eqref{sec:fit}, can also correctly extract the QNM amplitudes in this case. 

\begin{figure}[ht!]%
	\includegraphics[width=\columnwidth]{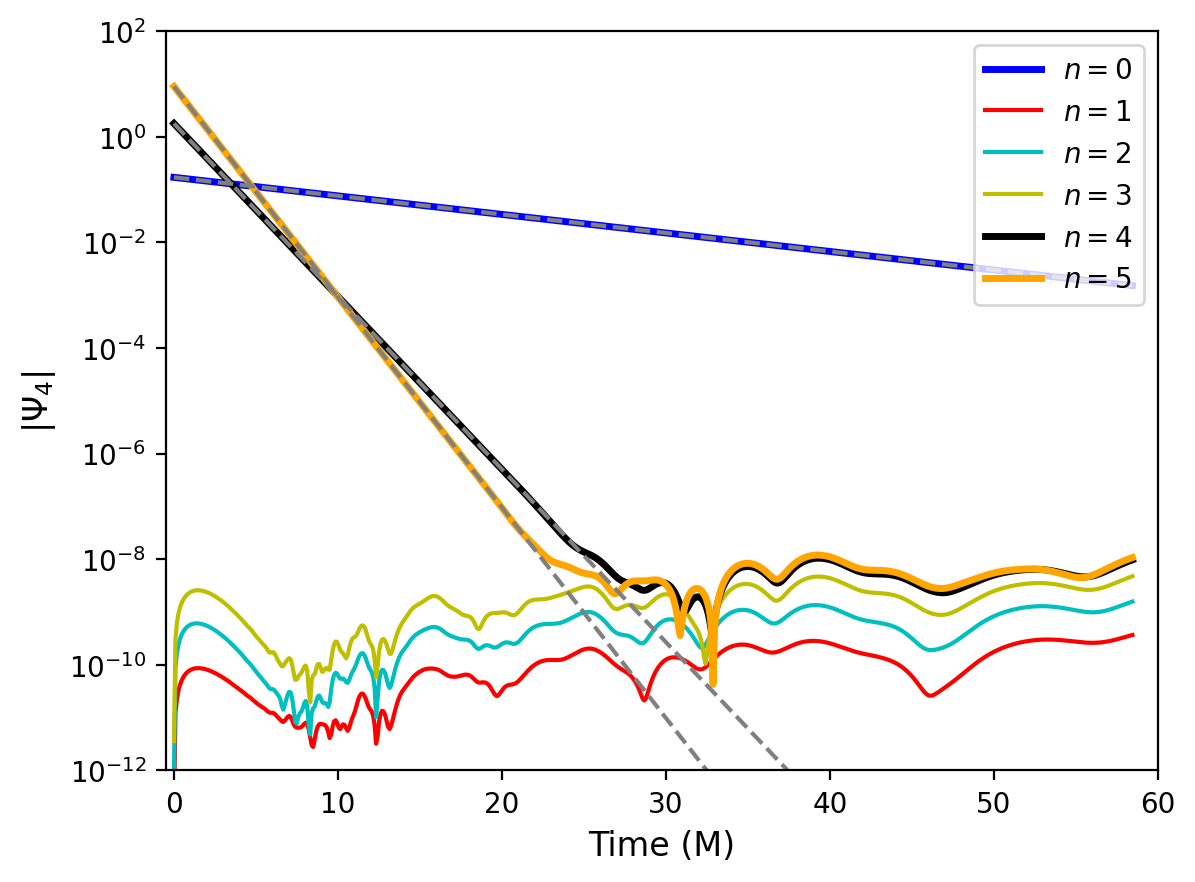}
  \caption{Extracted amplitudes when applying the spatial fit (colored solid lines), described in Sec.~\eqref{sec:exc} for an initial superposition of QNMs ($n=0$, $n=4$ and $n=5$) on a Kerr background with $a=0.7$. 
  Modes that are present in the initial data are plotted in bold lines. 
  The gray dashed lines show the expected mode amplitude given our QNM initial conditions. 
  The amplitudes for all modes are recovered to numerical precision, while the modes that are not present in the initial data have extracted amplitudes consistent with truncation error.}
  \label{fig:multi_mode}
\end{figure}

Let us consider initial data constructed by superposing the $l=m=2$ fundamental ($n=0$), fourth ($n=4$), and fifth ($n=5$) overtones on a Kerr background with spin $a=0.7$ (the expected remnant spin from the astrophysical merger of equal mass, non-spinning, quasi-circular binaries~\cite{Pretorius:2005gq}). 
In Fig.~\eqref{fig:multi_mode}, we show the amplitudes extracted by applying the spatial fit at different $t=$ constant surfaces (colored, solid lines) that match, up to numerical error, the analytical values of the mode amplitudes (gray, dashed lines). 
As a check, we have also included overtones that are \emph{not} present in the initial data in our fit, to demonstrate that the results have amplitudes consistent with the numerical error of our initial data and time evolution code. 
This test demonstrates the robustness of our fitting procedure, at least when applied to linearized solutions to the Einstein equations, with purely QNM initial data.

\begin{figure}%
	\includegraphics[width=\columnwidth]{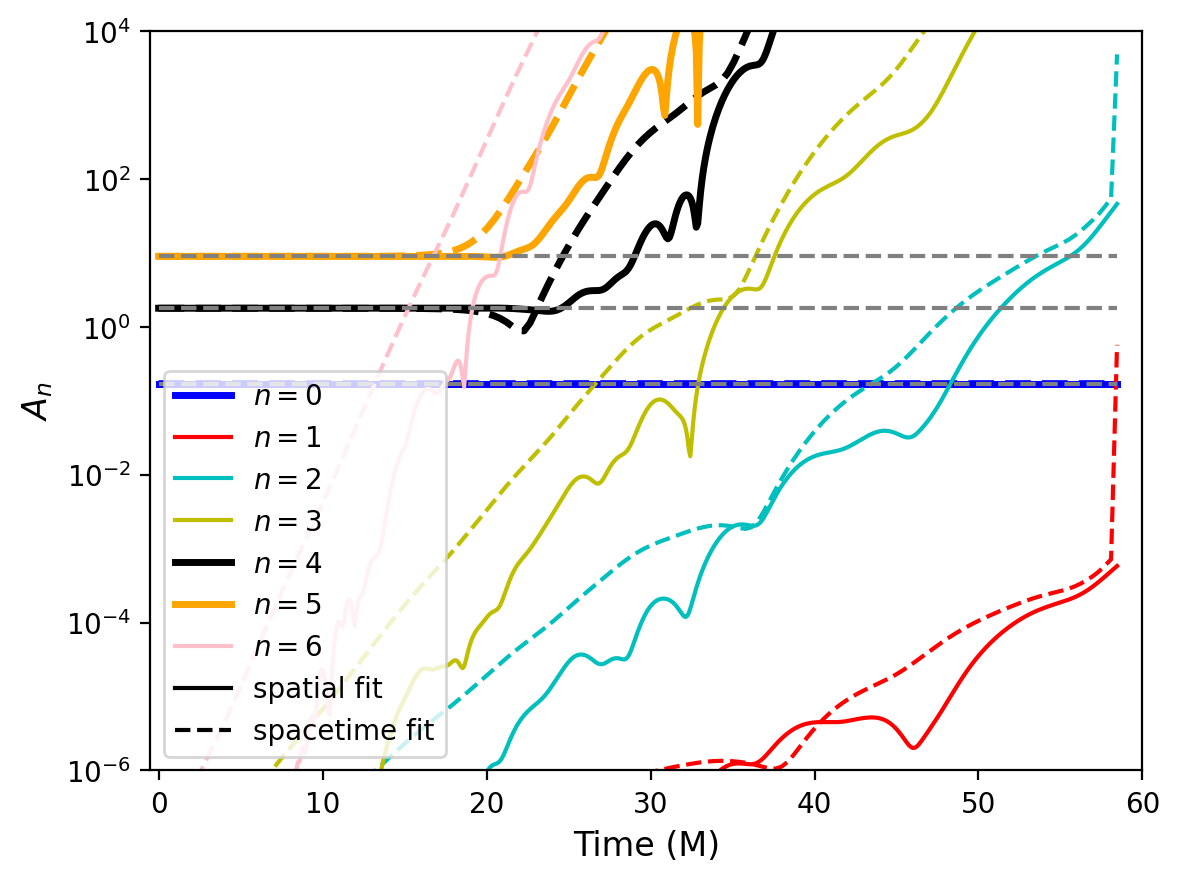}
  \caption{Stability of fitting superpositions of QNMs ($n=0$, $n=4$ and $n=5$) on a Kerr background with $a=0.7$ (from the same run as shown in Fig.~\eqref{fig:multi_mode}). 
  When factoring out the decay rate, the mode amplitudes we extract become constant in time, until the numerical noise dominates. 
  The amplitudes extracted from the spacetime fit (colored dashed lines) are consistent with those obtained from the spatial fitting (colored solid lines), and both agree with the analytical results (gray, dashed lines). 
  As expected, when fitting the overtones that were \emph{not} included in the initial data, the amplitudes are always unstable.}
  \label{fig:multi_mode_stab}
\end{figure}

Furthermore, as in, e.g., Ref.~\cite{Baibhav:2023clw}, in Fig.~\eqref{fig:multi_mode_stab} we show the stability of the fitting by factoring out the known decay rates of the modes. 
By doing that, the resulting QNM amplitudes are expected to be constant when fitting at different times, i.e., we consider the extraction of a given mode to be {\em stable} if we recover a roughly constant amplitude (and phase) over some relevant, non-negligible time period. 
We have also compared the results between the spatial fit (colored, solid lines) and the spacetime fit (colored, dashed lines). 
We find that both methods are capable of stably extracting all the QNMs present (even the fifth overtone) until their amplitudes reach a floor set by numerical truncation error. 
This suggests that the inner product presented in Sec.~\eqref{sec:exc} indeed establishes orthogonality between modes, complementing recent analytical results~\cite{Green:2022htq,London:2023prep}.

\subsection{Evolving and fitting to scattering initial data} \label{sec:scatter}

For our final example, we apply our quasinormal mode fitting procedures to analyze scattering initial data. 
These types of initial data excite the black hole in a more complex manner than quasinormal mode initial data, and we anticipate that a prompt, non-QNM transient solution to the Teukolsky equation will be noticeable in the ringdown signal. 
Specifically, we consider an approximately ingoing Gaussian pulse\footnote{Gaussian in Boyer--Lindquist radial coordinate $r_{BL}$.} as initial data:
\begin{subequations}
\label{eq:approx-ingoing-pulse}
\begin{align}
    \Psi_4(\rho,\theta) &= \exp\left\{ -\frac{(\rho^{-1}-r_0)^2}{w^2}\right\}~ _{-2}Y_l(\theta)\\
    \partial_t\Psi_4(\rho,\theta) &= - \frac{\rho^2}{2+4\rho} \partial_\rho(\Psi_4(\rho,\theta))~,
\end{align}
\end{subequations}
where $r_0$ and $w$ specify the initial central location and width of the pulse, respectively. 
For the context of this paper, regardless of the black hole spin, we specify the angular part of the initial data to be purely $l=m=2$ spin weight $-2$ spherical harmonics, and the radial part as a Gaussian centered at $r_0=8$M with width $w=1$. 
For Kerr black holes, we expect the $l>2$ modes to also be excited due to spherical-spheroidal mixing. 
To account for this mixing, we include up to $l=4$ modes when constructing the design matrices for fitting, and we include up to the $n=5$ overtones, both prograde and retrograde, for the $l=2$ modes and up to $n=1$ for $l=3,4$ modes, unless otherwise specified\footnote{We checked that the quality of the fit does not improve upon adding more modes, either higher harmonics or overtones.}. 

\begin{figure}%
	\includegraphics[width=\columnwidth]{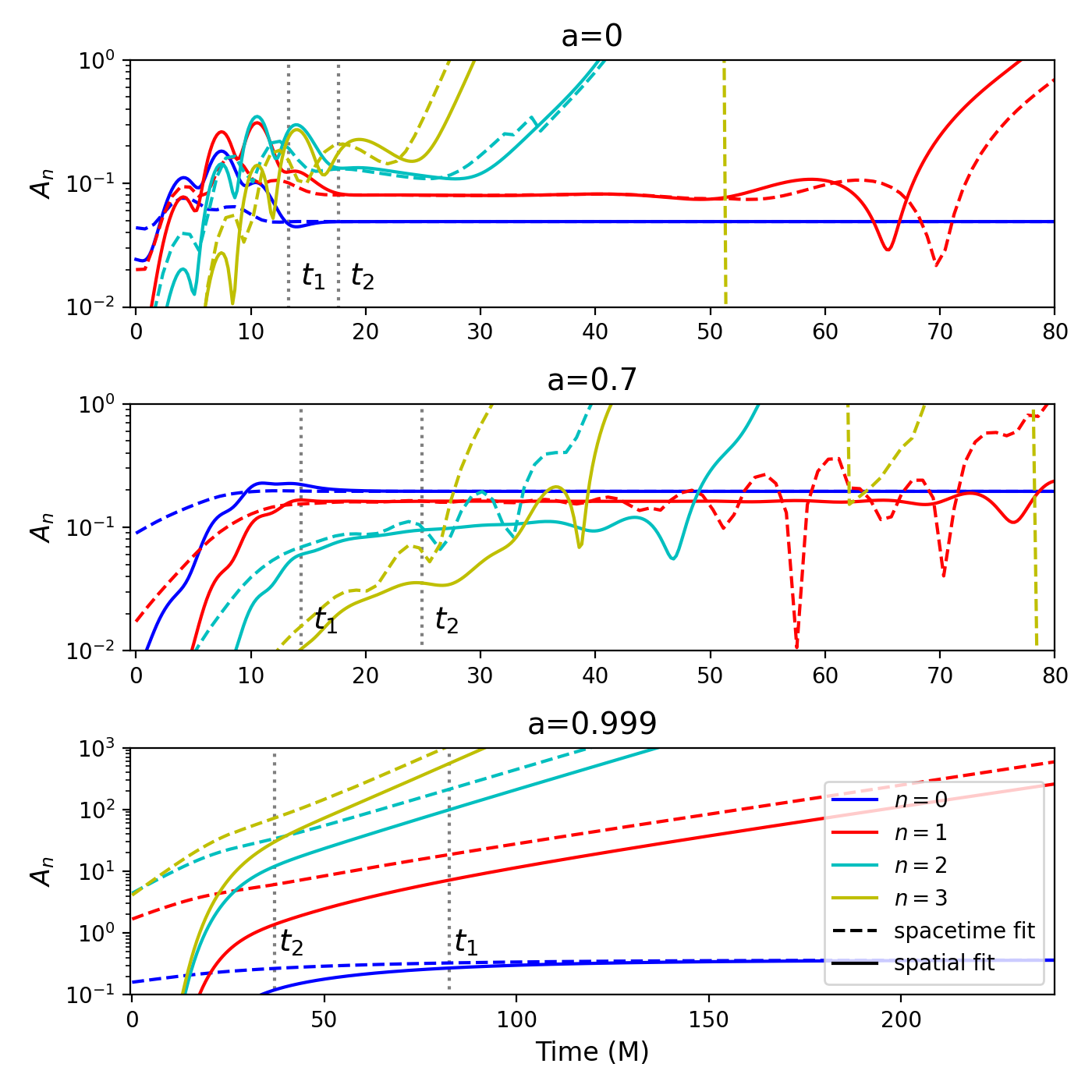}
  \caption{Extracted amplitudes with spatial fitting (solid lines) and spacetime fitting (dashed lines) from numerical scattering experiments on Kerr backgrounds with $a=0$ (top), $0.7$ (middle), and $0.999$ (bottom) as a function of the fitting start time. As in Fig.~\eqref{fig:multi_mode_stab}, the existence of a specific mode is supported if its amplitude remains constant over some period of time. The vertical dotted lines correspond to the times $t_1$ and $t_2$, when the amplitude of $\psi_4$ is maximum at the horizon and at infinity, respectively. We have scaled out the decay rate for the mode amplitudes relative to the value at $t_2$. Note that for the near-extremal case, $a=0.999$, the waveform first peaks at future null infinity ($t_2$) before peaking at the horizon ($t_1$); this is likely related to the transient growth of modes that in the extremal limit becomes the Aretakis instability~\cite{Aretakis:2012ei,Gralla:2016sxp}, which complicates the fitting for the overtones.}
  \label{fig:scattering}
\end{figure}

In Fig.~\eqref{fig:scattering}, we show the fitting results for different modes after applying the spatial fitting (solid lines) and spacetime fitting (dashed lines) procedures to the numerical data obtained with the scattering initial data on Kerr backgrounds with spins $a=0$, $a=0.7$ and, $a=0.999$.
To assess the stability of the fits over time, we anchored the amplitude of each mode at a common time $t_2$, chosen (somewhat arbitrarily) to be the time when $\Psi_4$ peaks at future null infinity; that is, we divide the mode amplitude at the time of fitting (the horizontal axis) by the expected amplitude evolution from the time $t_2$ to the fitting time. The subsequent fit is then stable if the fitted amplitude and phase remain constant over an interval of fitting start times.
For clarity, we have only plotted the overtones for the prograde, $l=m=2$ mode in Fig.~\eqref{fig:scattering}. The fundamental retrograde modes and higher multipoles from spherical-spheroidal mixing can also be extracted stably using our fitting methods.

Both fitting methods yield consistent results, although they inherently have different truncation error (or, loosely speaking, ``noise'') characteristics. We speculate that as the spacetime fitting method uses the late-time signal, its effective signal-to-noise ratio decreases faster than the spatial fit's method. Consequently, the mode amplitude computed from this method typically becomes unstable slightly earlier than that from the spatial fit. On the other hand, spatial fitting does not incorporate information from late times, hence it is more sensitive to the early-time transient, and, consequently, tends to become stable slightly later than the spacetime fitted amplitudes. 

Overall, for these scattering experiments we found that we can stably extract the fundamental mode and the first two overtones for a period of at least $\sim15$~M, around or after the time of peak $\psi_4$, for Kerr backgrounds not too close to extremal case
\footnote{The difficulty with near-extremal cases is not unexpected due to the slower decay of transients, related to the more efficient trapping of null geodesics. The instabilities in fitting could be due to insufficient integration time and larger numerical truncation error. We leave the detailed analysis of high spin cases to a future work.}. 
However, the fitting for higher overtones was generically unstable\footnote{In Appendix~\eqref{app:convergence} we show that the instability of overtones with $n\geq 3$ is not due to numerical error.}. 
Given that in the previous section we demonstrated that our code and fitting algorithms are capable of solving for and extracting superpositions of QNMs with overtones higher than the second, 
and that linear theory tells us the difference between these two classes of initial data resides in the transient part of the solution (as discussed above), this suggests that the source of the fitting instability is the presence of transients, with scattering initial data. 
We note, however, the $n\geq3$ overtones could be more strongly excited during mergers, and hence still be stably fitted. We defer the study of such initial data to a future work. 

\section{A comparison to the traditional time-only fitting at future null infinity} 
\label{sec:time-only}

In this section, we test the quality of a QNM fit from a traditional (time-only) fitting method, see, e.g. Ref.~\cite{Giesler:2019uxc}, and compare that against our fitting procedures.
The time-only fit we employ here is equivalent to our spacetime fit restricted at future null infinity; namely it takes into account the angular eigenfunction (spherical-spheroidal mixing) but without any radial information.
In numerical relativity, it is common to estimate the value the waveform takes at future null infinity, either through extrapolating waveforms measured at several finite radii, or through a Cauchy-characteristic extraction/matching~\cite{Bishop:1996gt,Reisswig:2009rx,Bishop:2016lgv,Barkett:2019uae,Moxon:2020gha,Moxon:2021gbv,Ma:2023qjn}, and then to find the best-fit QNMs using fits to the temporal evolution of a select set of angular harmonics of the waveform at infinity.

We will assess the quality of these temporal fits in two ways. First, we consider the stability of the fitting, namely how well one can stably extract the amplitude and phase when changing the fitting start time~\cite{Baibhav:2023clw}. Second, we consider the recovery of spatial information by testing whether the extracted mode amplitudes from performing time-only fits at several different radii agree with the radial eigenfunctions for the modes.

\subsection{Stability of the time-only fitting}

\begin{figure}%
	\includegraphics[width=\columnwidth]{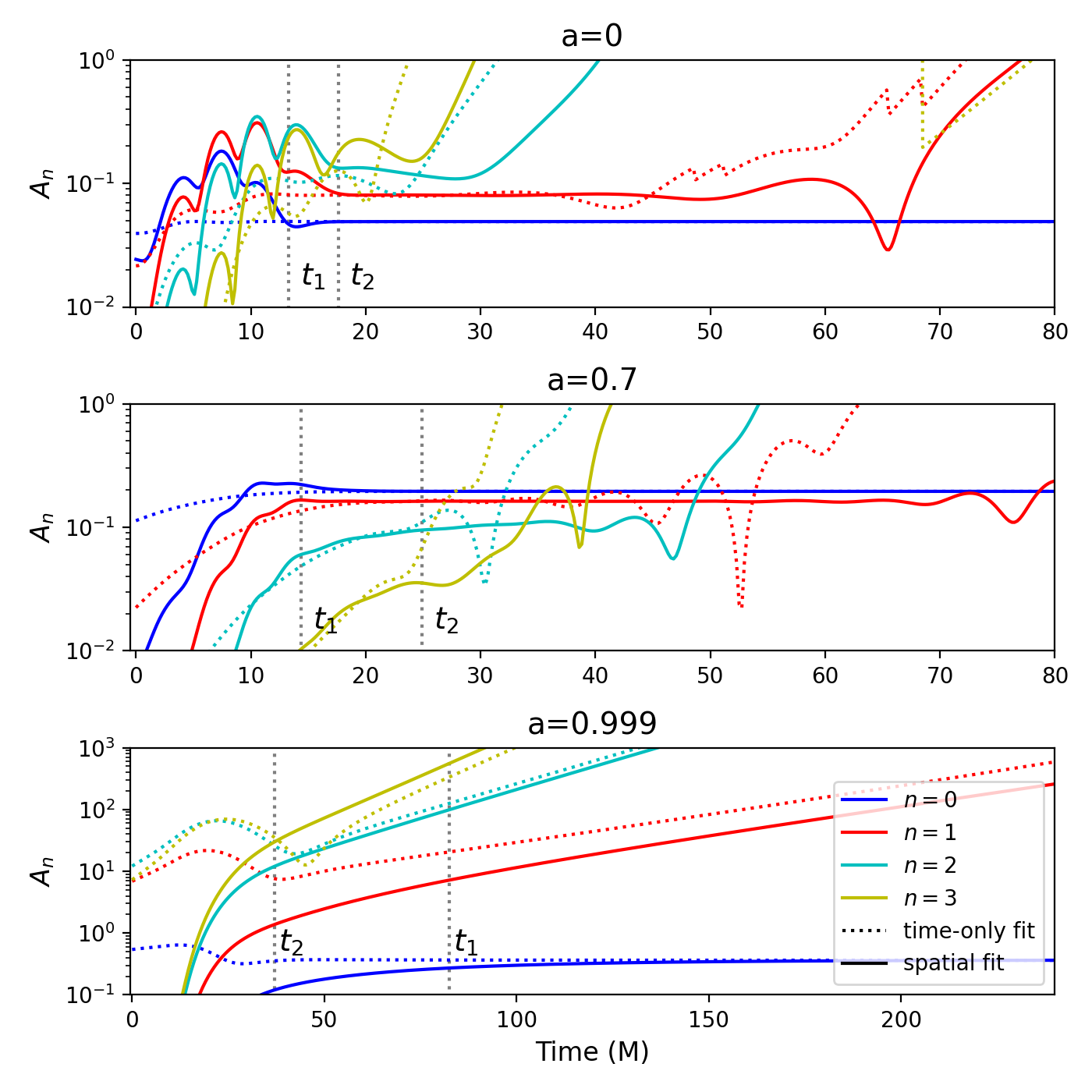}
  \caption{Comparison between the time-only fit (dotted lines) and the spatial fit (solid lines) applied to scattering initial data for a Kerr black hole with $a=0$ (top), $0.7$ (middle), and $0.999$ (bottom).} 
  \label{fig:qne_stability}
\end{figure}

In Fig.~\eqref{fig:qne_stability} we compare the results from the time-only fit at future null infinity (dotted lines) to the spatial fitting (solid lines), applied to the Gaussian pulse scattering experiment described in Sec.~\eqref{sec:scatter}. 
We find that, for Schwarzschild and Kerr with $a=0.7$, the time interval over which the first overtone's ($n=1$) amplitude and phase can be stably extracted is much shorter in duration with the time-only fit, while the second and higher overtones can never be stably extracted with the time-only fit\footnote{This quantitative result may not hold true for merger calculations, as one would expect the relative excitation amplitudes for the overtones in mergers to be different from that in the scattering initial data presented here. Though we do expect that qualitatively a similar result would hold, namely that above a certain overtone number $n$ QNMs cannot be stably measured due to a combination of transients and numerical error.}. 
Interestingly, we find that when applying the time-only fit to the $a=0.7$ Kerr black hole, the first overtone can be stably extracted \textit{before} $\Psi_4$ peaks at future null infinity. 
Whether the above holds in astrophysical ringdown (that is, with initial data that smoothly match to the gravitational wave signal after merger) needs further study, but earlier results do indicate that, at least with numerical relativity waveforms one can decompose the signal into QNMs beginning at the time of peak strain \cite{Buonanno:2006ui,Giesler:2019uxc}, roughly ~$10$ M before the peak of $\Psi_4$ (issues of overfitting aside).

The time-only fit's improved stability at early times in Fig.~\eqref{fig:qne_stability} does not indicate superior performance to the spatial fit. As the time-only fit makes use of the signal over a time interval, at early times it still attempts to fit the post-peak QNM ringing. This results in a smooth rise in the mode amplitudes for the time-only fit with respect to the fitting start time, until they asymptote to stable values. By contrast the spatial fit uses the information at a constant time slice. We interpret the instabilities in the mode amplitudes at early times for the spatial fit as a demonstration of the fact that the QNMs are not a complete description of generic black hole ringdown signals

\subsection{Recovery of the Spatial Eigenfunction}

Another robust way to assess if the waveform is being overfitted (or underfitted) is to test the time-only QNM fits at different radii against the known radial eigenfunction.
Starting with the mode contribution of the signal, Eq.~\eqref{eq:mode}, if one fits it with the QNM frequencies at a set of radii, one would expect the measured amplitude for each mode labeled by $l$, $n$, and $[p]$ to be radially scaled by the associated radial wave function:
\begin{equation} \label{eq:time_amp}
    A_{[p]ln}^{t}(\rho) = R_{[p]ln}(\rho) A_{[p]ln}~,
\end{equation}
where $R_{[p]ln}(\rho)$ is the radial quasinormal eigenfunction, $A_{[p]ln}$ is the true amplitude of the quasinormal mode, and $A_{[p]ln}^{t}(\rho)$ is the expected amplitude one gets from a time-only fit at a given radius $\rho$. 

In the presence of a transient contribution to the ringdown signal, the time-only fit can be susceptible to overfitting. 
When overfitting, the measured radial structure of modes from the time-only fit would deviate from Eq.~\eqref{eq:time_amp} as transients have distinct radial functions from QNMs.
To quantify the agreement between the time-only fit for a QNM, labeled by subscript $n$, and the radial eigenfunction, we define the following spatial mismatch $\mathcal{M}$:

\begin{equation}\label{eq:mismatch}
    \mathcal{M} = 1-\left\vert\frac{\langle A^{t}_{n}(\rho),R_n(\rho)\rangle}{\sqrt{\langle A^{t}_{n}(\rho),A^{t}_{n}(\rho)\rangle\langle R_n(\rho),R_n(\rho)\rangle}}\right\vert~,
\end{equation}
where
\begin{equation}
    \langle f,g\rangle = \int d\rho \bar{f}g~.
\end{equation}

\begin{figure}%
	\includegraphics[width=\columnwidth]{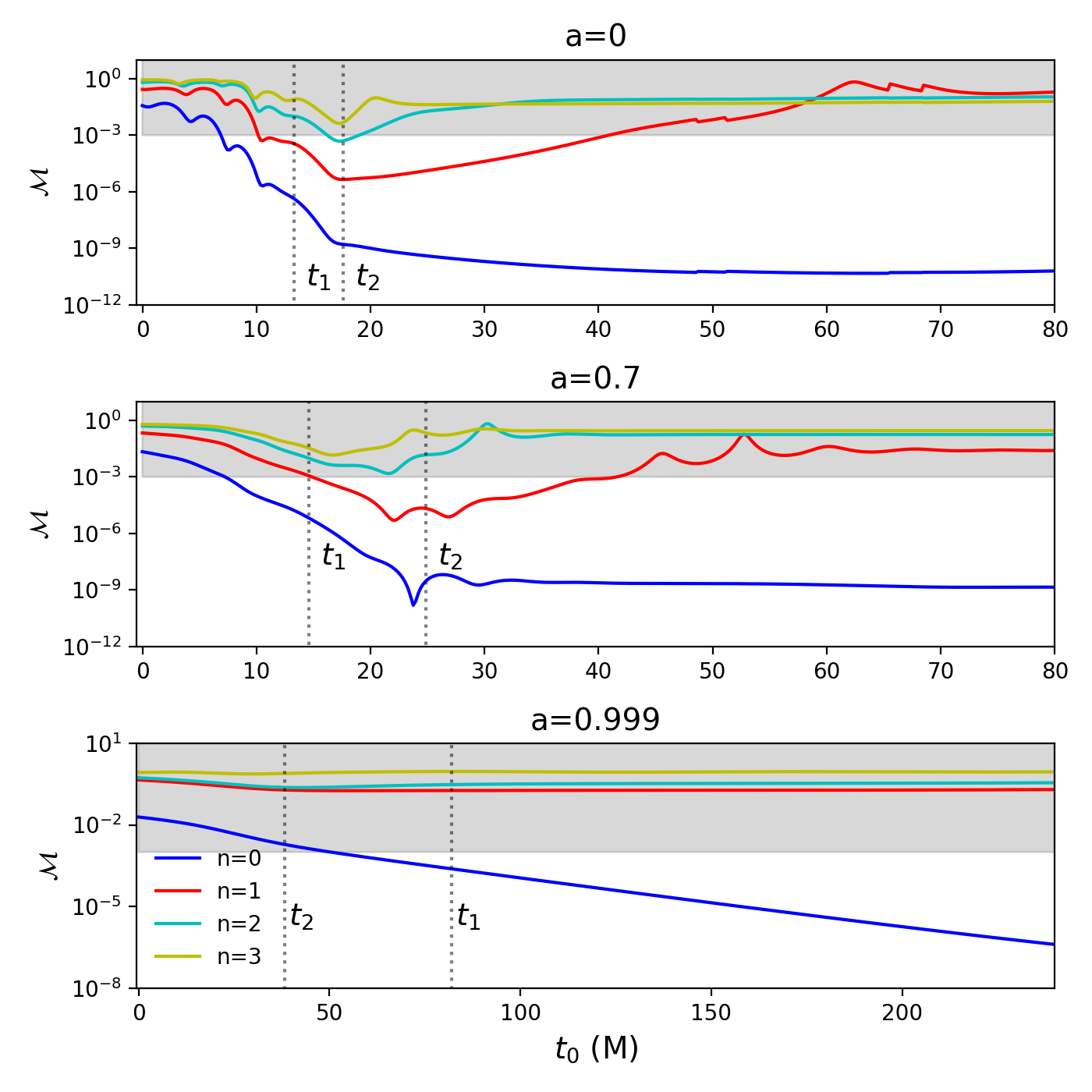}
  \caption{Spatial mismatch $\mathcal{M}$ (Eq.~\ref{eq:mismatch}) for the time-only fit as a function of fitting start time $t_0$, applied to scattering experiments for a Kerr black hole with $a=0$ (top), $a=0.7$ (middle), and $a=0.999$ (bottom). The radial amplitude variation agrees relatively well with the known radial function when $\mathcal{M} < 10^{-3}$, i.e., outside the shaded region.} 
  \label{fig:mismatch}
\end{figure}
We now perform a waveform fit at several different fixed radii on numerical data with the scattering initial data described in Sec.~\eqref{sec:scatter}, and test if the fitted amplitude and phase for each mode at different radii agree with the prediction from the radial eigenfunction. 
We vary the fitting start time $t_0$ for the time-only fit and evaluate the radial mismatch $\mathcal{M}$ as a function of $t_0$. The result is shown in Fig.~\eqref{fig:mismatch}.
We find that for Schwarzschild, (with scattering initial data), the time-only fit can identify the fundamental mode and first two overtones. 
For Kerr with spin $a=0.7$, we can only faithfully reconstruct up to the first overtone\footnote{The marginal dip in the mismatch for the $n=2$ mode around $t_0=15\sim 20M$ hints at its existence for Kerr with $a=0.7$, which we do extract stably using the spatial fit (see Fig.~\ref{fig:qne_stability}).}. 
Finally, in the near-extremal limit, the radial mismatch for the fundamental mode decreases as $t_0$ increases because the transient decays away faster than the fundamental mode, yet none of the overtones are correctly recovered within the time span of our numerical integration ($500$ M). 

\begin{figure}
	\includegraphics[width=\columnwidth]{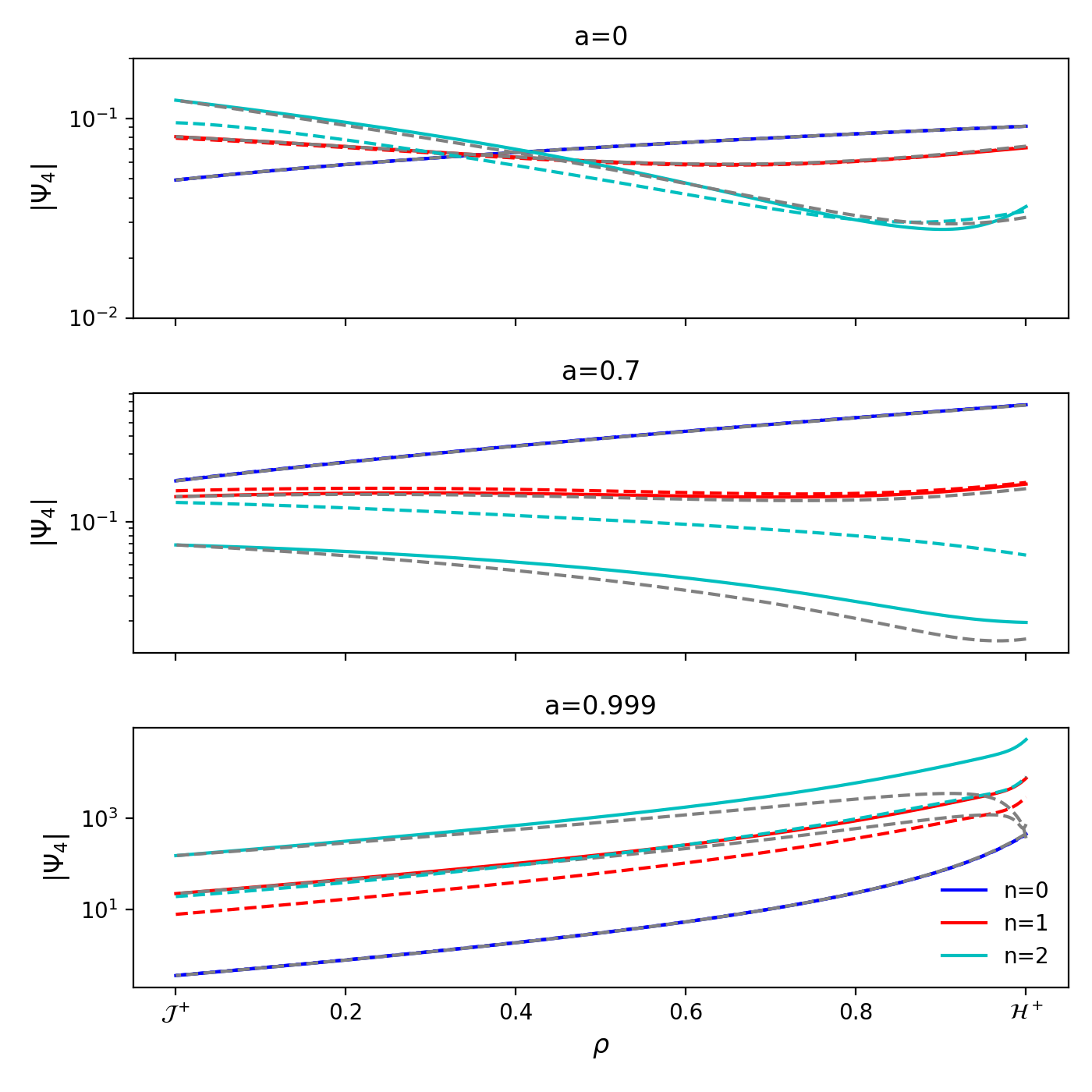}
  \caption{Mode amplitude radial variation from time-only fitting (colored lines) for scattering initial data (see Eq.~\eqref{eq:approx-ingoing-pulse}). 
  Note that the vertical axis for each subplot has a different scale. 
  We plot the measured radial amplitude variation with two fitting start times $t_1$ (colored, solid lines) and $t_2$ (colored, dashed lines), the time at which the waveform peaks at $\mathcal{H}^+$ and $\mathcal{J}^+$ respectively. 
  The known radial functions are plotted as gray, dashed lines for comparison, whose amplitudes are chosen to match the solid colored lines at $\mathcal{J}^+$. 
  The seemingly better agreement for the overtones in the case of $a=0.999$ near null infinity is likely due to the similar frequencies of the family of zero-damped modes in the extremal limit, i.e., approaching this limit the overtones do not decay much more rapidly that the fundamental mode. }
  \label{fig:qne_recov}
\end{figure}

To illustrate the quality of the reconstructed radial structure, in Fig.~\eqref{fig:qne_recov}, we plot the mode amplitudes as a function of radius from time-only fits against the expected radial eigenfunctions (gray, dashed lines), with $t_0 = t_1$ (colored, solid lines) and $t_2$ (colored, dashed lines), the times at which the waveform peaks at the horizon and null infinity respectively. 
For visual comparison, the amplitudes for the known radial functions are set to agree with the time-only fit at future null infinity when $t_0=t_1$ (solid lines), by construction. 

As indicated by Figs.~\eqref{fig:mismatch} and~\eqref{fig:qne_recov}, the radial eigenfunctions are better recovered by the time-only fit at a surprisingly \emph{early} time (overfitting aside), except for the near-extremal case. 
We further note that our ability to extract the QNM radial variation through the time-only fit also depends on the initial data, which, as we already discussed, heavily impacts the form of the transient signal. 
We defer a detailed study of the initial data and interpretation of the seemingly better-behaved fittings at early time to a future work. 

\section{Discussion}\label{sec:conclude}

In this work we have presented two new techniques for extracting the quasinormal mode content from perturbed single black holes. 
The main novel aspect of the fitting procedures is they utilize the radial structure of each QNM over the full exterior of the black hole spacetime. 
This is aided by our use of horizon penetrating, hyperboloidally compactified coordinates, in which the quasinormal mode eigenfunctions to the Teukolsky equation are well behaved from the horizon to future null infinity.

We used the methods described in Refs.~\cite{Ripley:2022ypi,Ripley:2020xby} to solve the Teukolsky equation in the time domain, evolving initial data that can be a superposition of a chosen number of QNMs together with a more generic transient. 
We first showed that our fitting procedures are capable of stably extracting the correct amplitudes when the initial data consist of a superposition of pure QNMs, and include rapidly decaying high overtones, until the corresponding amplitudes drop below a floor set by numerical truncation error. 
The reason that the fitting procedure works this well is that it uses more information about the waveform -- namely, its radial dependence. 
This drastically increases the number of data points available in the fit as compared to a fit at a fixed radius (or at a small number of radii). Moreover, by making use of the radial dependence of the modes, we can construct an inner product for which the quasinormal modes are orthogonal with respect to each other. This allows us to project out QNM amplitudes from a perturbation consisting of a pure sum of QNMs at a given fixed time slice, though such projection can still be biased in the presence of transients (defined as any non-QNM component of the perturbation). 

With confidence in our ability to accurately extract QNMs in the absence of transients, we examined the linear excitation of a black hole through a prompt, compact pulse of gravitational wave. This investigation aimed to shed light on the issue of {\em overfitting} when attempting to extract the excitation amplitudes of QNMs from numerical simulations of binary black hole mergers, which can lead to erroneous QNM amplitude measurements if transients are not accounted for. As we are using the Teukolsky equation for the evolution of the perturbation, we can only study the effects of the linear transients. However, since it seems unlikely that non-linear effects would help with the problem of overfitting, our study can be considered a best-case scenario for the theoretical measurement of excitation amplitudes. 
First, we showed that even using our new fitting algorithms, the presence of linear transients (with scattering initial data) prevented stable measurement beyond the second overtone of the fundamental mode of the dominant $\ell=m=2$ perturbation for a Schwarzschild and an $a=0.7$ Kerr black hole (for an $a=0.999$ Kerr black hole we could only stably extract the fundamental mode during our integration time of $500$M). 

We then compared our new fitting procedures to a more traditional time-only fit. 
This analysis showed that a time-only fit may result in erroneous amplitude for the first overtone of the $\ell=m=2$ mode of an $a=0.7$ Kerr black hole, outside an interval for fitting start time of order $15$ M.
Moreover, when performing the time-only fit at different radii, we found that the amplitude and phase one obtains from the fitting at each radius does not match the predicted behavior of the second (and higher) overtones of the quasinormal mode radial eigenfunctions (again, except for the case of Schwarzschild where the second overtone does match to reasonable precision). 
In the case of a near-extremal hole ($a=0.999$), only the fundamental mode can be faithfully extracted due to long-lasting transient instability near the horizon; the overtones might be present in the signal but require longer time of integration and higher numerical resolution.

A significant issue regarding extrapolating our results to what this implies for existing studies, which have attempted to extract mode amplitudes in the full non-linear case, is that we have not attempted to match our Gaussian pulse perturbation of the black hole to the initial data of any particular merger event, as we do not have theoretical estimate for the excitation amplitudes of the overtones in merger case. 
Thus, though we expect similar issues to occur in the non-linear case at some overtone number $n$ for any given angular mode, and expect overfitting to be worse in the non-linear case, to put a threshold on the cutoff overtone number from the linear problem would require a study using better adapted initial data. 

The basic idea of the techniques described here--fitting for the spatial behavior of the quasinormal eigenfunctions and their dependence on time--could, in principle, be applied to fully nonlinear numerical relativity simulations of ringdown. 
However, there are several complications: the method requires a fixed gauge choice in the ringdown phase (potentially achievable through an appropriate gauge driver condition~\cite{Lindblom:2007xw,Lindblom:2009tu}), and a careful treatment of wave extraction in the strong field~\cite{Zhang:2012ky}. 
Additionally, spatial information could only be used far away from the remnant black hole, where the gravitational waves could be well described by linear (and possibly second-order) perturbation theory.

Our ability to set up pure mode initial data should allow for further studies of nonlinear, second-order effects during black hole ringdown. Studying the time evolution of pure quasinormal mode initial data in a nonlinear/second order code would allow one to systematically study the efficiency of mode mixing.
Additionally, with pure quasinormal mode eigenfunction initial data, one could study the functional form of the second order source term that appears in the solution to the second-order vacuum Teukolsky equation~\cite{Loutrel:2020wbw,Ripley:2020xby}. Doing so would allow us to study how the source term varies with different kinds of quasinormal modes, such as the overtones. We leave a study of these effects to future work.

Our setup---solutions to the Teukolsky equation with fitting procedures that use the entire waveform, not just its value at future null infinity---is arguably ``optimal'' for extracting the QNM signal. 
Specifically, the solutions we studied do not exhibit nonlinearities, allowing us to concentrate solely on the signal's prompt and QNM contributions. 
Do astrophysical mergers excite the overtones of the remnant black hole more cleanly compared to the scattering experiments proposed here? We leave this question to a future study to set up initial data describing the perturbed remnant black hole from merger calculations. 
Nevertheless, given the challenges of fitting for the overtones in this (simplified) setup, our results provide further evidence that fitting for the overtones in astrophysical or full numerical relativity data, as well as the interpretation thereof, is a highly sensitive process that depends significantly on the data extraction and fitting procedures employed.

\section*{Acknowledgments}

We thank Emanuele Berti, Will Farr, Elena Giorgi, Abhishek Hegade, Stefan Hollands, Lam Hui, Maximiliano Isi, Macarena Lagos, Lionel London, Nicholas Loutrel, Sizheng Ma, Keefe Mitman, Rob Owen, Harrison Siegel, and Zihan Zhou for their useful comments regarding various aspects and physical implications of this project. 
H.Z. especially thanks Will Farr, Maximiliano Isi, Sizheng Ma, and Harrison Siegel for discussions regarding fitting procedures. 
We also thank an anonymous referee for their constructive comments.
The authors are pleased to acknowledge that the work reported on in this paper was substantially performed using the Princeton Research Computing resources at Princeton University which is consortium of groups led by the Princeton Institute for Computational Science and Engineering (PICSciE) and Office of Information Technology's Research Computing.
J.L.R. acknowledges support from the Simons Foundation through Award number 896696 and the National Science Foundation through award PHY-2207650. A.C.-A. acknowledges support from the Simons Foundation.

\appendix
\section{Calculating quasinormal modes and eigenfunctions for the Teukolsky equation}\label{app:qne}

As we discuss in Sec.~\eqref{sec:numerical-results}, one set of initial data we use consists of a linear superposition of quasinormal eigenfunctions (QNEs).
To compute the quasinormal modes, we follow the algorithm presented in \cite{Ripley:2022ypi}, except for one change that we found allows us to stably solve for the higher overtones. 
Here we briefly outline the algorithm, and the improvement to it (the code we used can be accessed at \cite{justin_ripley_2023_8215611}).

The Teukolsky equation \eqref{eq:hyperboloidal_teuk_eqn} separates under the following decomposition
\begin{align}
    \Psi\left(\tau,r,\theta,\phi\right)
    =
    e^{im\phi - i\omega \tau}R\left(\rho\right)S\left(\theta\right)
    .
\end{align}
From this, we obtain two ordinary differential equations, which we schematically write as
\begin{widetext}
\begin{subequations}
\label{eq:radial-swal-eqn}
\begin{align}
    \label{eq:radial-eqn}
    A\left(\rho\right)\frac{d^2R}{d\rho^2}
    +
    B\left(\omega,m,\rho\right)\frac{dR}{d\rho}
    +
    \left(C\left(\omega,m,\rho\right)-{}_s\Lambda^m_{l}\right)R
    &=
    0
    ,\\
    \label{eq:swal-eqn}
    \frac{1}{\sin\theta}\frac{d}{d\theta}\left(\sin\theta \frac{dS}{d\theta}\right)
    +
    \left(s + \frac{\left(m+s\cos\theta\right)^2}{\sin^2\theta} - 2a\omega s\cos\theta + a^2\omega^2\cos^2\theta + {}_s\Lambda^m_{l}\right)S
    &=
    0
    ,
\end{align}
\end{subequations}
\end{widetext}
where $A,B,C$ are functions that are given in \cite{Ripley:2022ypi}. Note that \eqref{eq:swal-eqn} is the standard equation for the spin-weighted spheroidal harmonics \cite{Teukolsky:1973ha}. 
Following \cite{Ripley:2022ypi}, we view \eqref{eq:radial-swal-eqn} as defining two eigenvalue problems with the eigenvalue ${}_s\Lambda^m_{l}$. 
The set of $\left\{\omega,R,S\right\}$ for which \eqref{eq:radial-eqn} and \eqref{eq:swal-eqn} have the same eigenvalue ${}_s\Lambda^m_{l}$ are the quasinormal modes and eigenfunctions of the Teukolsky equation \cite{Ripley:2022ypi}. We note that \eqref{eq:radial-eqn} and \eqref{eq:swal-eqn} also admit total transmission and scattering mode solutions, but they would be irregular at the outer boundary. Therefore, the regularity one inherits from the set of spectral basis functions eliminates such solutions.

We numerically discretize \eqref{eq:radial-eqn} and \eqref{eq:swal-eqn}, solve for the eigenvalues and eigenvectors of the two systems, and then vary $\omega$ until at least one of the eigenvalues for the two discretized systems coincide.
The value $\omega$ is then a quasinormal mode frequency, and the corresponding eigenvector with eigenvalue ${}_s\Lambda^m_l$ gives the quasinormal eigenfunction. 
As in \cite{Ripley:2022ypi} we discretize \eqref{eq:swal-eqn} using a spectral method first described in \cite{Hughes:1999bq,Cook:2014cta}.
The radial equation \eqref{eq:radial-eqn} was discretized using a Chebyshev pseudospectral method in \cite{Ripley:2022ypi}.
We found that solving for the higher overtone quasinormal modes using the radial Chebyshev pseudospectral method required using a large number of collocation points. 
This led to numerically ill-conditioned discretizations of \eqref{eq:radial-eqn}, which then required the use of higher-precision arithmetic.
Here we describe a spectral method that leads to sparse, well-conditioned discretizations of \eqref{eq:radial-eqn}, even when we solve for the higher quasinormal mode overtones. 

The spectral method makes use of the properties of the ultraspherical (also called the Gegenbauer) polynomials \cite{olver2013fast}.
For completeness, we outline the basic idea of the method here, although we refer to \cite{olver2013fast} for a more complete exposition.
Ultimately we used the \texttt{ApproxFun} \cite{ApproxFun.jl-2014}\footnote{\url{https://github.com/JuliaApproximation/ApproxFun.jl}} implementation of these methods in our code.
Our conventions follow \cite{NIST:DLMF}. 

We first transform the radial coordinate to $x\in\left[-1,1\right]$.
We next expand $R$ in a series of Chebyshev polynomials of the first kind
\begin{align}
    \label{eq:summation-R-rep}
    R(x)
    =
    \sum_{n=0}^N c_n T_n(x)
    .
\end{align}
The derivative of the Chebyshev polynomials of the first kind can be written
in terms of the Chebyshev polynomials of the second kind
\begin{align}
    \frac{d T_n(x)}{dx}
    =
    n U_n(x)
    .
\end{align}
For higher-order derivatives with respect to $x$, we use the following property of the ultraspherical polynomials:
\begin{align}
    \frac{d C^{(\lambda)}_n(x)}{dx}
    =
    2\lambda C^{(\lambda+1)}_{n-1}(x)
    ,
\end{align}
where $U_n(x) = C^{(1)}_n(x)$.
To conclude, we see that we can write
\begin{subequations}
\label{eq:differentiation-ultraspherical}
\begin{align}
    \frac{dR}{dx}
    =
    \sum_{n=1} n c_n U_{n-1}(x)
    .\\
    \frac{d^2R}{dx^2}
    =
    \sum_{n=2} 2nc_n C_{n-2}^{(2)}(x)
    .
\end{align}
\end{subequations}
Consider the vectorial representation of $R$, Eq.~\eqref{eq:summation-R-rep}
\begin{align}
    R
    =
    {\bf T}^T {\bf c}
    ,
\end{align}
where ${\bf T} = \left(T_0\left(x\right),T_1\left(x\right),...,T_N\left(x\right)\right)$ and ${\bf c}\equiv \left(c_0,c_1,...,c_N\right)$.
We see that we can write the first and second derivatives of $R$ as
\begin{subequations}
\begin{align}
    \frac{dR}{dx}
    &=
    {\bf U}^T\mathbb{D}_1 {\bf c}
    ,\\
    \frac{d^2R}{dx^2}
    &=
    {\bf C}^T\mathbb{D}_2 {\bf c}
    ,
\end{align}
\end{subequations}
where ${\bf U} = \left(U_0\left(x\right),U_1\left(x\right),...,U_N\left(x\right)\right)$, ${\bf C}\equiv \left(C^{(2)}_0\left(x\right),C^{(2)}_1\left(x\right),...,C^{(2)}_N\left(x\right)\right)$, and $\mathbb{D}_1$ and $\mathbb{D}_2$ are sparse matrices, the components of which can be inferred from Eqs.~\eqref{eq:differentiation-ultraspherical}.
To complete the discretization of \eqref{eq:radial-eqn}, we need to convert $A,B,C$, along with $dR/d\rho$ and $R$ to the polynomial basis $C^{(2)}_n$, which can be done using sparse matrices \cite{olver2013fast}.
Ultimately with this method, \eqref{eq:radial-eqn} can be discretized to take the form
\begin{align}
\label{eq:generalized_eigen_problem}
    \left(
        \mathbb{A} 
        - 
        \lambda \mathbb{B}
    \right)
    \textbf{c} 
    = 
    0
    .
\end{align}
where $\mathbb{A}$ and $\mathbb{B}$ are sparse matrices with a relatively small condition number.
We do not need to impose boundary conditions as regularity at the boundaries imposes the ingoing radiation condition at the black hole horizon and the outgoing radiation condition at future null infinity \cite{zenginouglu2008hyperboloidal,Ripley:2022ypi}.
We solve the generalized eigenvalue problem \eqref{eq:generalized_eigen_problem} using standard numerical linear algebra routines (that is, with the \texttt{eigen} solver in the \texttt{Julia} standard library).

\section{Structure of QNM Radial Eigenfunctions}\label{app:qnm_radial}
Here, we briefly discuss the structure of the radial eigenfunctions for QNMs on $\tau=const.$ HPHC hypersurfaces. 
In HPHC coordinates, $\tau=const.$ hypersurfaces become tangent to null surfaces at the future horizon and null infinity (and hence tangent to the characteristic curves of the Teukolsky equation).
There are two main effects that determine the far-field behavior of the radial quasinormal eigenfunctions to the Teukolsky equation.
First, there is some flexibility in HPHC coordinates as to where $\tau=const.$ hypersurfaces intersect future null infinity (while pinned to the same location at the horizon) \cite{PanossoMacedo:2019npm}; we call this flexibility a \emph{propagation effect}. 
Second, the rate at which the coordinate volume on the slice changes as a function of $\rho$ controls the behavior of the eigenfunction at future null infinity, 
which gives rise to the familiar 1/r decay at large radii. 
Since we solve for the rescaled variable $\Psi_4$ defined in Eq.~\eqref{eq:Psi4} in both the QNM (initial data) code and the evolution code, the $1/r$ volume effect is factored out.
We discuss the propagation effect in detail below.
\subsection*{Propagation effects}

\begin{figure}
   \centering
   \begin{tikzpicture}
       /pgfplots/y=0.1cm, /pgfplots/x=0.1mm 
      \draw [very thick] 
         (0,0) -- (2.5,2.5) -- (0,5) -- (-2.5,2.5) -- (0,0);
      \draw [very thick] 
         (2.6,2.6) -- (3.1,3.1) -- (0.6,5.6) -- (0.1,5.1) -- (2.6,2.6);
      \draw [black,dashed, very thick] 
         (-2,3) .. controls (-1,2.2) and (1.2,2) .. (2,3);
     \begin{axis}[
        anchor=origin, 
        rotate around={135:(current axis.origin)}, 
        xmin=0., ymin=-0.05, ymax = 0.05, xmax = 3, clip=false,
        axis lines=none, 
        xtick=\empty, ytick=\empty,
        ]
    \addplot [thick, blue, domain=0:1.5, smooth] {0.005*sin(x*1000)*exp(-0.5*x)-0.0705};
    \addplot [thick, red, domain=0:1.5, smooth] {0.005*sin(x*1000)*exp(-2*x)-0.0705};

    \end{axis}
      \draw [gray, dotted ,very thick] 
         (-0.5,0.5) -- (2,3);
      \draw [gray, dotted ,very thick] 
         (-1.5,1.5) -- (1,4);
      \node at (2.5,4.3) {$\mathcal{J}^+$};
      \node at (-1.6,4) {$\mathcal{H}^+$};
      \node at (0.05,5.3) {$i^+$};
      \node at (2.8,2.5) {$i^0$};
      \node at (-2.8,2.5) {$\mathcal{B}$};
   \end{tikzpicture}
   \caption{Penrose diagram of the Kerr exterior. 
   The black dashed curve describes a $T=const.$ hypersurface in HPHC coordinates.
   The gray dotted lines represent the trajectories of outgoing null geodesics.  
   The box in the top-right corner shows the observed black hole ringdown signal at future null infinity, one with slower decay (blue) and the other faster (red). 
   Note that the signals are illustrative only; any decaying wave would have infinite amplitude at $i^0$, zero amplitude at $i^+$, and infinitely many oscillations between.}
\label{fig:penrose_diagram_bh}
\end{figure}
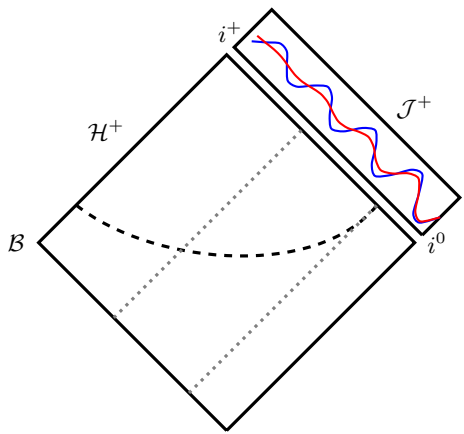

To understand the nature of propagation effects in HPHC coordinates, we first solve the null geodesic equation in ingoing Eddington--Finkelstein coordinates (for a related discussion see Appendix C of \cite{Ripley:2020xby}). 
Setting $\xi_\theta=\xi_\phi=0$, we find that the characteristic speeds of out- and ingoing null geodesics are
\begin{subequations}
\begin{align}
    c_+ &= \frac{\xi^+_v}{\xi^+_r} = 1-\frac{4Mr}{2Mr+\Sigma_{BL}} \\
    c_- &= \frac{\xi^-_v}{\xi^-_r} =  -1.
\end{align}
\end{subequations}
To determine the radial null characteristics on a hyperboloidal slice, we first define a radial coordinate $\rho(r)$ and time coordinate $T(v,r)$. 
Under that coordinate change the characteristic speeds are
\begin{align}
    \tilde{c}_\pm &=  \frac{d\rho/dr}{\frac{1}{c_{\pm}}\partial_v T + \partial_r T}.
\end{align}
From this, we can determine the time that it takes for a radially outgoing wave, starting at radius $\rho_0$ to reach null infinity by integrating 
\begin{equation}\label{eq:plus_time} 
    \tau_+(\rho_0) 
    = 
    \int^{\rho_{\mathcal{J}^+}}_{\rho_0}\frac{1}{\tilde{c}_+(\rho)}d\rho.
\end{equation}
Similarly, one can compute the time in this coordinate for a radially ingoing wave to reach the black hole horizon:
\begin{equation} \label{eq:minus_time}
    \tau_-(\rho_0) = \int_{\rho_{\mathcal{H}^+}}^{\rho_0}\frac{1}{\tilde{c}_-(\rho)}d\rho~,
\end{equation}
where $\rho_\mathcal{J^+}$ and $\rho_{\mathcal{H}^+}$ are, respectively, the radius of null infinity and horizon in this coordinate which we assume to be independent of time. 

One may interpret these time intervals as the amount by which the slice mismatches with a radially out-/ingoing spherical wave front. 
As the mode is exponentially decaying, the amplitude of the mode will be affected by this time mismatch.
For a quasinormal mode with frequency $\omega$, the amplitude variation of the radial wave function due to this mismatch time, for the outgoing part of the wave, is given by:
\begin{equation}\label{eq:amp_pre}
    A(\rho)\propto \exp\{\Im\{\omega\} \tau_+(\rho)\}.
\end{equation}
Note that here the amplitude increases faster toward infinity for a wave with a higher decay rate. 

Figure~\eqref{fig:penrose_diagram_bh} diagrams the basic intuition behind this result.  
Far from the black hole, a spherical wave would be approximately advected with a decay of $1/r$ along the null geodesics. 
After factoring out the $1/r$ decay, we expect the amplitude of the wave to be roughly constant along the outgoing null geodesics labeled by grey dashed lines at large radius. 
We see that on a $T=const.$ hypersurface, the faster decaying mode would have a radial amplitude $A(\rho)$ that decays faster as we approach the black hole. 

\begin{figure}%
	\includegraphics[width=\columnwidth]{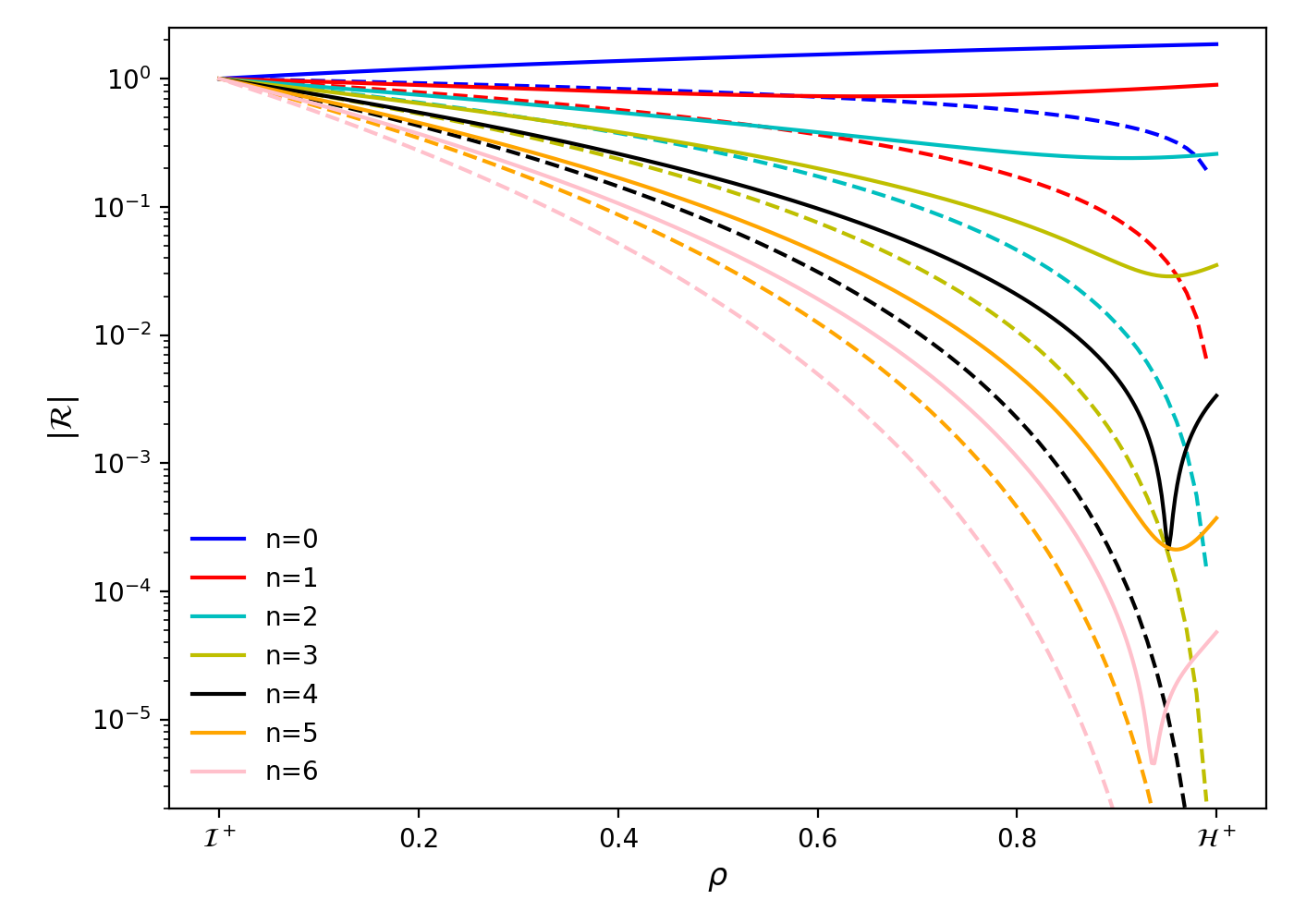}
  \caption{Radial eigenfunctions of overtones for Schwarzschild with $l=m=2$ (solid, colored lines). We also plot predictions from radially outgoing geodesics with dashed lines, by evaluating Eq.~\eqref{eq:amp_pre} with the overtone frequencies. 
  We note that the geodesic prediction yields larger slopes for the radial functions at $\mathcal{J}^{+}$ for overtones with higher n, in accordance with the eigenfunctions we calculated from the Teukolsky equation. 
  However, the slope for radial eigenfunctions is not precisely matched by the geodesic prediction.
  This is likely due to the fact that quasinormal modes do not exactly satisfy the eikonal limit from which \eqref{eq:amp_pre} is derived.}
  \label{fig:qne_s}
\end{figure}

We note that the propagation time depends on the geodesic trajectory one considers (although the natural choice is to compute the propagation time from the characteristic speed of the Teukolsky equation, as we do here).
Since QNMs carry angular momentum, we expect that the propagation times \eqref{eq:plus_time} and \eqref{eq:minus_time} are lower bounds on the propagation time of a quasinormal mode wavefront. 
We also note that the wavefront of a mode only travels along a null geodesic in the eikonal limit; finite-wavelength effects may further complicate our above argument. 
Nevertheless, we show that the QNM radial functions roughly follow the above argument in Fig.~\eqref{fig:qne_s}. 
In that figure we plot the radial eigenfunction for the overtones following the procedure outlined in Appendix~\eqref{app:qne}, and compare them to the predicted scaling of the radial amplitude from \eqref{eq:amp_pre}. 
In particular, we present the radial eigenfunctions for successive overtones of a Schwarzschild black hole with $l=m=2$. 
We see that the radial profiles of the modes roughly follow the scaling as predicted by Eq.~\eqref{eq:amp_pre}.

To determine $\tau_+\left(\rho\right)$, we note that in the coordinates we are using
\begin{align}
    \rho(r) &= \frac{L^2}{r} \\
    T(v,r) &= v + h(r)~, 
\end{align}
where
\begin{equation}
    \frac{dh}{dr} = -1 - \frac{4M}{r}~.
\end{equation}
Here, $L$ is a constant length scale that we take to be 1. The locations of the horizon and future null infinity are then $\rho = \frac{1}{M+\sqrt{M^2-a^2}}$ and $\rho=0$, respectively.
The in- and outgoing characteristic speeds in these coordinates are
\begin{align}
    \tilde{c}_+ &= -\frac{a^2 \rho^2 \cos (\theta )-2 M
   \rho +1}{8 M^2-4 a^2 M \rho  \cos
   (\theta )} \\
   \tilde{c}_- &= \frac{\rho ^2}{4 M \rho +2}~.
\end{align}
For illustrative purposes we calculate the propagation time defined in Eqs~\eqref{eq:plus_time} and~\eqref{eq:minus_time}, for null rays for a $M = 1/2$, $a=0$ black hole. 
In this case, the in- and outgoing propagation times become:
\begin{subequations}
\begin{align}
    \tau_+ (\rho) &= 2 \log \left(1 -\rho\right)\\
    \tau_- (\rho) &= 2 - \frac{2}{\rho}+2 \log\left(\rho\right)~.
\end{align}
\end{subequations}
Note that the outgoing time delay diverges at the horizon, and the ingoing time delay diverges at infinity; this reflects the fact that there is no outgoing radiation at the black hole horizon and no ingoing radiation at future null infinity.

\section{Convergence Tests}\label{app:convergence}

In this section we show the numerical convergence of time domain code~\cite{justin_ripley_2023_8302934} used in this work.
We consider the evolution of a single QNM, and a numerical scattering experiment. 
The time evolution code makes use of pseudo-spectral methods in the angular ($\theta$) direction and fourth order finite difference methods in the radial $(\rho)$ direction. 

The initial data are then integrated in time using an explicit fourth-order Runge--Kutta integrator. 
Therefore, fixing the angular resolution, one expects the code to approach the continuum solution with fourth order convergence. 
In general, we find that the numerical error is dominated by the radial direction. 

\begin{figure}%
	\includegraphics[width=\columnwidth]{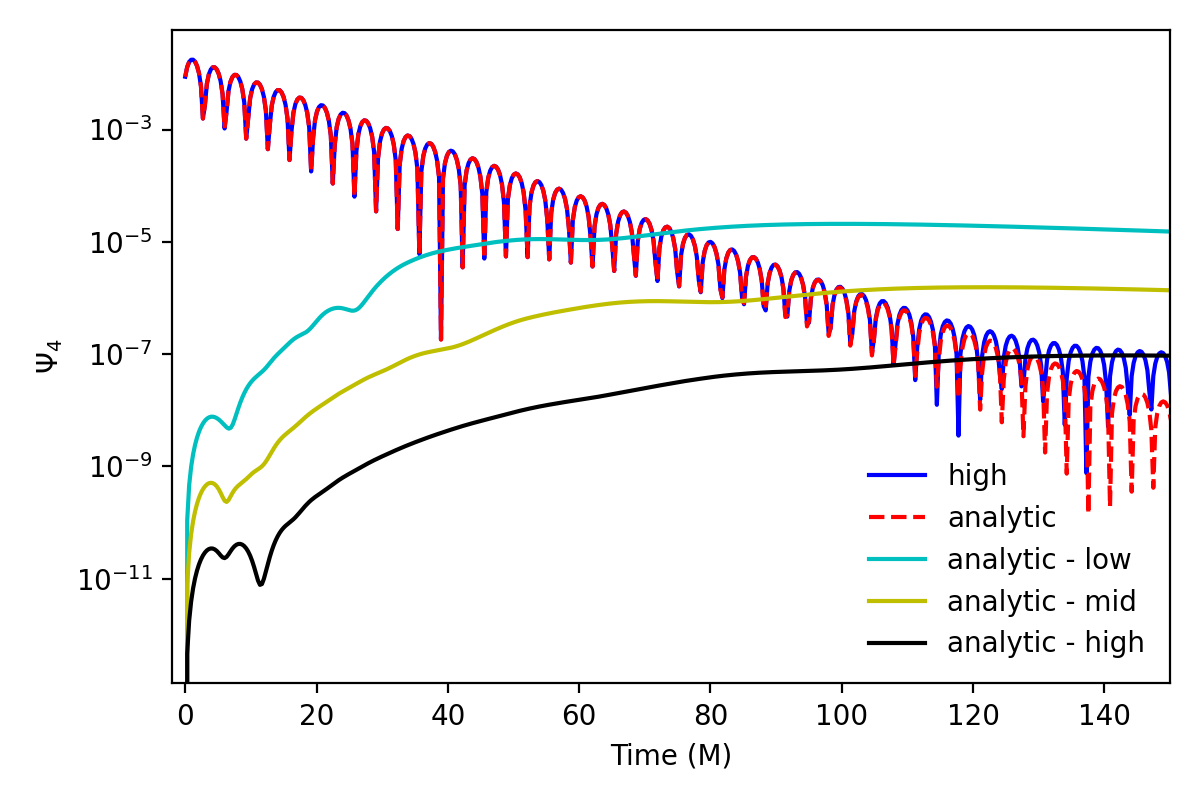}
  \caption{Convergence evolving a single QNM with $n=3$ and $a=0.999$. Here, the number of angular collocation points is fixed to $40$ for all runs, and we change the radial resolution by factors of $2$ successively. The low resolution is run with $n_\rho=128$, mid with $n_\rho=256$, and high with $n_\rho=512$ grid points. By the \emph{analytic} answer we mean the prediction for $\Psi_4$ at future null infinity for an $n=3$, $l=m=2$ quasinormal mode.}
  \label{fig:qnm_converge}
\end{figure}

\begin{figure}%
	\includegraphics[width=\columnwidth]{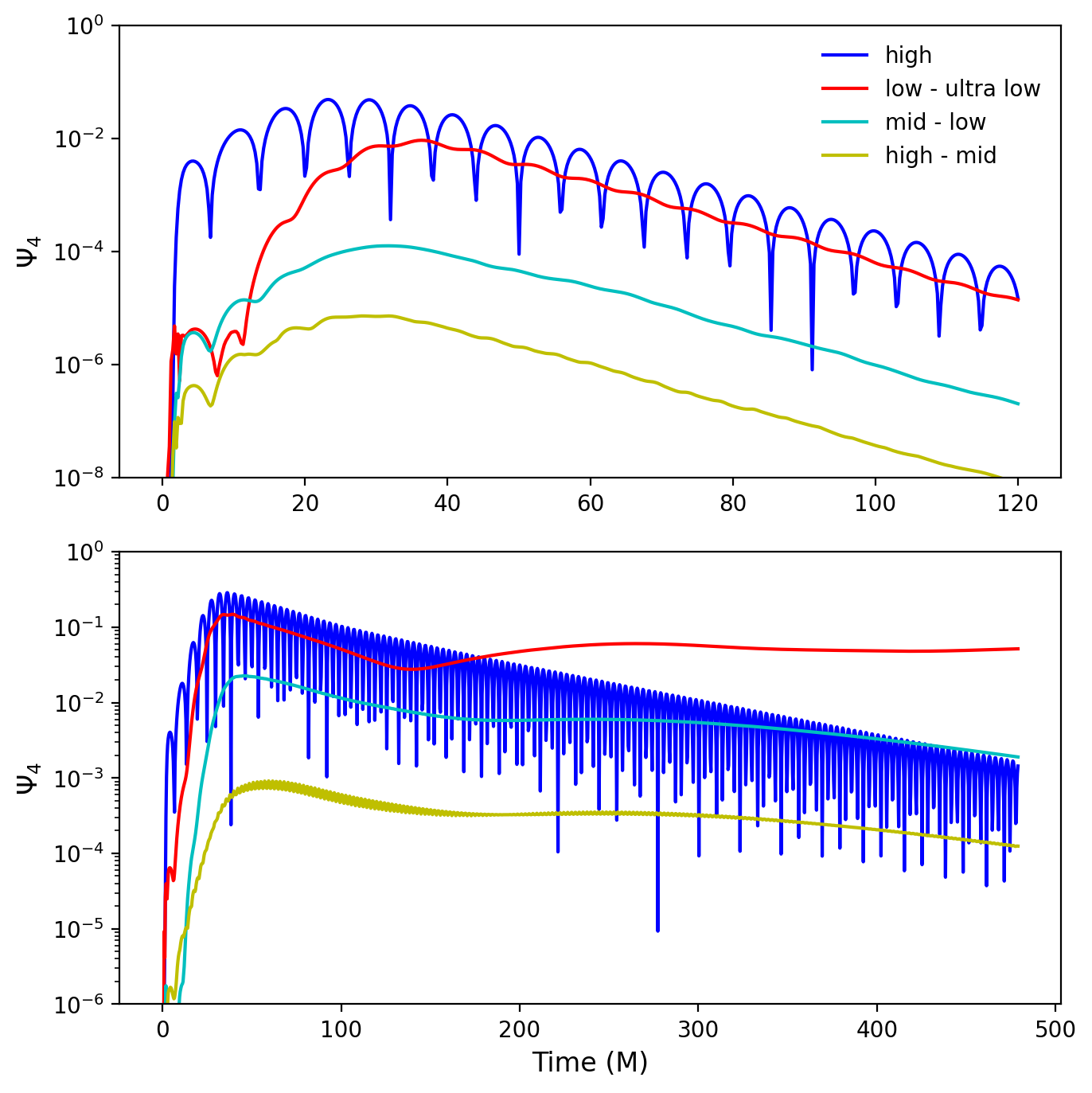}
  \caption{Convergence in a scattering experiment simulation off a Kerr background with $a=0.7$ (top) and $a=0.999$ (bottom). Here, the number of angular collocation points is fixed to $n_\theta =40$ for all runs, and we change the radial resolution by factors of $2$ successively. The ultralow resolution is run with $n_\rho=64$, low with $n_\rho=128$, mid with $n_\rho=256$, and high with $n_\rho=512$.}
  \label{fig:scattering_converge}
\end{figure}

\begin{figure}%
	\includegraphics[width=\columnwidth]{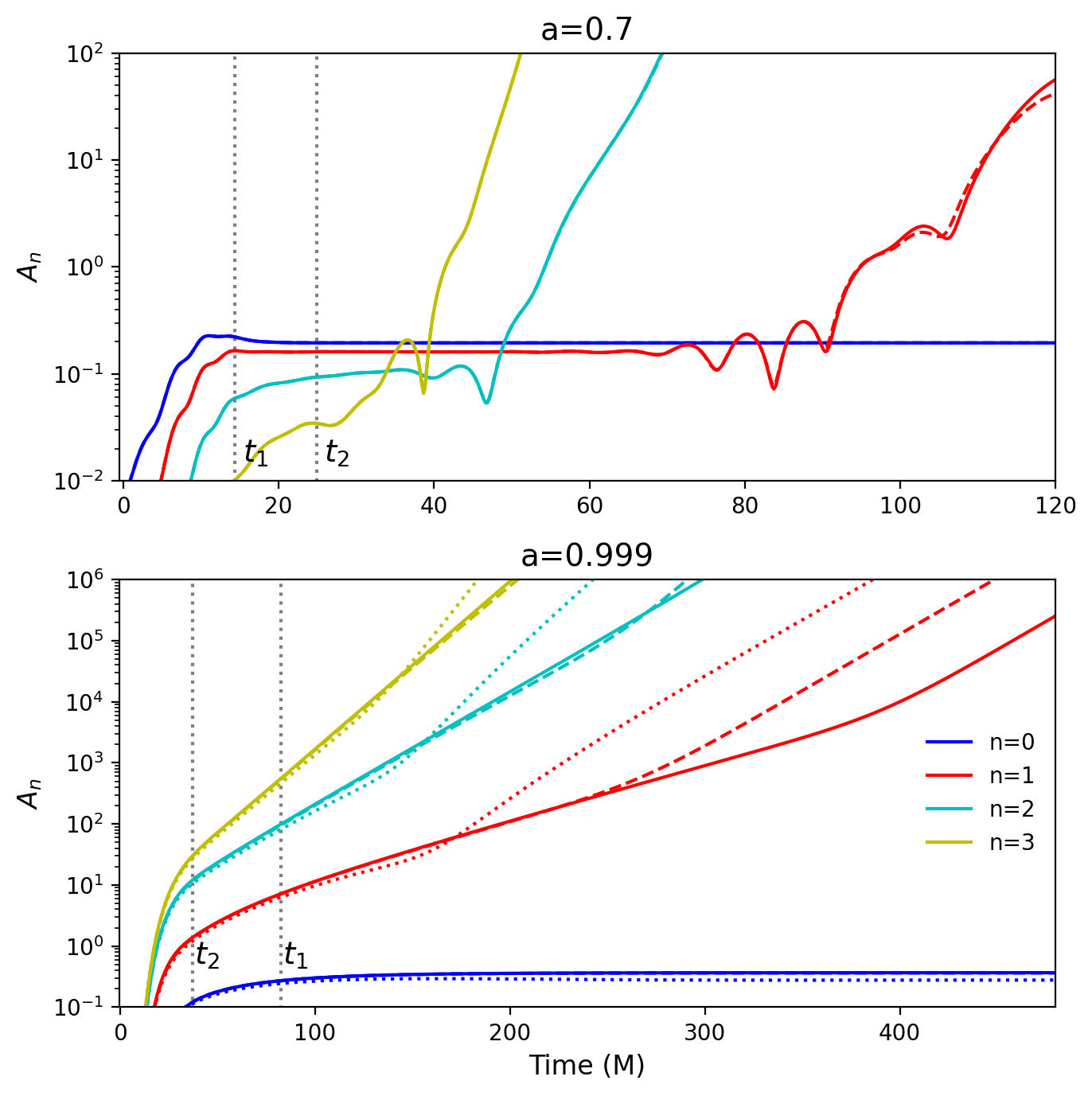}
  \caption{Convergence of fitting for the scattering initial data on a Kerr background with $a=0.7$ (top) and $a=0.999$ (bottom), here illustrated with the spatial fitting algorithm (spacetime fitting shows similar convergence properties). The solid lines show the fitting result from the high resolution simulation ($n_\rho=512$), and dashed lines show that from the midresolution ($n_\rho=256$); for $a=0.999$, we also plot the low resolution ($n_\rho=128$) in dotted lines. We find that for $a=0.7$ increasing numerical resolution does not improve the stability of fitting the overtones, indicating the source of instability is due to a resolved transient. For $a=0.999$, we find that the late-time kink in the slope for the first overtone (red lines) is due to the numerical truncation error; the kink moves to a later time as resolution improves, from around $180$M for the low resolution (red, dashed line) to around $400$M to the high resolution (red, solid line). However,  the transient dominates the first overtone  for at least the first $250$M of evolution, during which the mid- and high resolution agree.}
  \label{fig:fitting_convergence}
\end{figure}

For our convergence tests, we fix the number of angular collocation points to be $40$, and increase the radial resolution by successive factors of 2. 
We see fourth order convergence in Fig. \eqref{fig:qnm_converge}, for single QNM evolution, and in Fig. \eqref{fig:scattering_converge}, for the gravitational wave scattering experiment. We show that the numerical resolution of our simulations is not the limiting factor in the precision of our QNM fits to scattering initial data in Fig. \eqref{fig:fitting_convergence}, where we compare the spatial fit applied to both the high and midresolution runs. 

\clearpage
\bibliography{thebib}
\end{document}